\documentclass[twocolumn]{aastex62}
\usepackage{natbib}
\usepackage{apjfonts}
\usepackage{amsmath}
\usepackage{appendix}

\DeclareSymbolFont{cmletters}{OML}{cmm}{m}{it}
\DeclareMathSymbol{v}{\mathalpha}{cmletters}{"76}

\newcommand{\aeq}{\ensuremath{a_{\rm eq}}}
\newcommand{\msun}{\ensuremath{M_{\odot}}}

\graphicspath{{./}{figures/}}

\received{\today}
\revised{xxx}
\accepted{xxx}
\submitjournal{ApJ}
\shorttitle{Rapid BH spin-down by Thick MADs}
\shortauthors{Lowell et al.}

\begin{document}

\title{Rapid Black Hole Spin-down by Thick Magnetically Arrested Disks\footnote{Released on February, 1st, 2023}}

\correspondingauthor{Beverly Lowell}
\email{beverlylowell@u.northwestern.edu}

\author[0000-0002-2875-4934]{Beverly Lowell}
\affil{Center for Interdisciplinary Exploration $\&$ Research in Astrophysics (CIERA), Physics and Astronomy, Northwestern University, Evanston, IL 60202, USA}

\author{Jonatan Jacquemin-Ide}
\affiliation{Center for Interdisciplinary Exploration $\&$ Research in Astrophysics (CIERA), Physics and Astronomy, Northwestern University, Evanston, IL 60202, USA}

\author{Alexander Tchekhovskoy}
\affiliation{Center for Interdisciplinary Exploration $\&$ Research in Astrophysics (CIERA), Physics and Astronomy, Northwestern University, Evanston, IL 60202, USA}

\author{Alex Duncan}
\affiliation{Cornell University, Ithaca, NY 14850, USA}

\begin{abstract}

Black hole (BH) spin can play an important role in galaxy evolution by controlling the amount of energy and momentum ejected from near the BH into the surroundings. We focus on radiatively-inefficient and geometrically-thick magnetically-arrested disks (MADs) that can launch strong BH-powered jets. With an appropriately chosen adiabatic index, these systems can describe either the low-luminosity or highly super-Eddington BH accretion regimes. Using a suite of 3D general relativistic magnetohydrodynamic (GRMHD) simulations, we find that for any initial spin, a MAD rapidly spins down the BH to the equilibrium spin of $0< \aeq \lesssim 0.1$, very low compared to $\aeq = 1$ for the standard thin luminous (Novikov-Thorne) disks. This implies that rapidly accreting (super-Eddington) BHs fed by MADs tend to lose most of their rotational energy to magnetized relativistic outflows. In a MAD, a BH only needs to accrete 20\% of its own mass to spin down from $a=1$ to $a=0.2$. We construct a semi-analytic model of BH spin evolution in MADs by taking into account the torques on the BH due to both the hydrodynamic disk and electromagnetic jet components, and find that the low value of $\aeq$ is due to both the jets slowing down the BH rotation and the disk losing a large fraction of its angular momentum to outflows. Our results have crucial implications for how BH spins evolve in active galaxies and other systems such as collapsars, where BH spin-down timescale can be short enough to significantly affect the evolution of gamma-ray emitting BH-powered jets.

\end{abstract}


\section{Introduction} \label{sec:intro}

It is well established that supermassive BHs (SMBHs) ranging in mass from $10^6$ to $10^9 \msun $ lie at the center of nearly every galaxy. Active galactic nuclei (AGN) are  powered by the accretion of gas onto SMBHs. AGN release energy in the form of radiation, winds, and under the right circumstances, relativistic jets. The result is AGN feedback, and it is thought to play a key role in galaxy evolution \citep{2007ApJ...665.1038C, 2013ARA&A..51..511K}. Scaling relations between SMBHs and properties of their host galaxies (e.g. stellar velocity dispersion, \citealt{2000ApJ...539L...9F, 2003ApJ...596L..27K, 2021MNRAS.502L...1K}; bulge mass, \citealt{2003BHmassBulgemassVIRIAL,2004BHmassBulgemassJEANSEQN}; and total stellar mass, \citealt{2015ApJ...813...82R}) are powerful tools to understand the coevolution of BHs and galaxies. 

Massive jets heat and ionize the gas in and around galaxies, preventing cooling and stifling star formation, and sweep gas out of the galactic bulge \citep{1998A&A...331L...1S, 2005Natur.433..604D}. Moreover, the energy deposited in the surrounding environment affects the fueling of the BH itself, consequently regulating its duty-cycle.

The outflows are likely powered in part by the accretion of gas onto the SMBH. The Eddington limit is an important reference point in understanding BH growth, and thus galaxy structure and evolution. Estimations of the Eddington ratio from observations are largely uncertain. Most AGN accrete below the Eddington luminosity ($L_{\rm Edd}$), but there is evidence that some accrete at super-Eddington rates. In the super-Eddington regime, the flow becomes radiation-dominated, and the accretion disk is geometrically-thick. While many details of this accretion regime remain unclear, super-Eddington accretion is expected, and even inevitable, in some circumstances. Super-Eddington accretion can be observed in relatively local AGN \citep{lanzuisi_nustar_2016, tsai_superEdd_2018} and tidal disruption events \citep[TDEs,][]{1988Natur.333..523R}; it can be inferred in ultraluminous X-ray sources \citep[ULXs,][]{sutton_ulx_2013, kitaki_ulx_2017, 2018MNRAS.473.4360W}  and gamma-ray bursts \citep[GRBs,][]{1994ApJ...427..708P}. 

Whereas in low redshift AGN super-Eddington accretion episodes are rare, the conditions for super-Eddington accretion were favorable at higher redshifts \citep{aird_superEdd_2018}. The gas content of early galaxies correlates with the accretion rate of the SMBH, implying that a BH will accrete above the Eddington rate given sufficient gas supply \citep{2018PASJ...70L...2I}. Furthermore, there is no theoretical evidence that super-Eddington accretion is not possible \citep{2004cbhg.sympE..60S,2012ApJS..201....9F,2014MNRAS.439..503S,2019ApJ...880...67J}. 

Quasars at up to $z \approx 7$ have been found to have BH masses of $10^9 \msun$ \citep{2011Natur.474..616M, 2021ApJ...923..262Y}. To grow to such a large mass at that redshift from a reasonable BH seed, a BH would need to spend at least some time accreting at or above the Eddington rate \citep{2005ApJ...633..624V, 2015ApJ...804..148V, 2017MNRAS.464.1102B}. In fact, super-Eddington accretion episodes may be the most reasonable way to achieve $10^9 \msun$ SMBHs by $z \approx 7$, as maintaining accretion rates at or below the Eddington rate at the highest redshifts might require tuned conditions, although models that grow $M_{\rm BH}=10^9M_\odot$ SMBHs by $z=9$ through mergers of $10^4$ stellar-mass BHs, which formed at $z\sim20$ and had accreting most of their life at an Eddington rate since their formation, have been constructed \citep{Yoo2004}. More recently, James Webb Space Telescope has revealed SMBHs already at redshift $z=10.1$ \citep{2023arXiv230515458B}, challenging the models of SMBH formation and growth \citep{2023ApJ...955L..24G}, and suggesting either the existence of massive BH seeds or highly super-Eddington accretion at such high redshifts.

For a given luminosity, the lower the BH mass, the less gas reservoir is needed to achieve the Eddington rate. It follows that high accretion rates will more likely be found in galaxies hosting low-mass SMBHs. Cosmological simulations from \cite{2017MNRAS.472L.109A} showed that underweight BHs (those that lie below the $M_{\rm BH} - M_{\rm bulge}$ relation) go through super-Eddington accretion episodes until the bulge reaches a critical mass, and the accretion rate then levels off and the BHs join the correlation. The short timescales involved with these episodes make them difficult to be observed. Similarly, the tightly-correlated $M-\sigma$ relation may not represent all local AGN BH masses, but is instead an upper bound on BH mass and represents BHs accreting at or below the Eddington rate \citep{King2010}. If that is the case, BHs that lie below the $M-\sigma$ relation transiently accrete at super-Eddington rates: such outbursts may last a rather short time and thus are difficult to catch in the act. Even if we were to catch one of these BHs feasting on gas at a super-Eddington rate, the observed luminosity would be likely limited to just a few Eddington luminosities due to the low radiative efficiency of super-critical accretion \citep{1988ApJ...332..646A}.

An example of this may be the Narrow line Seyfert 1 (NLS1) galaxies, which are thought to host central BHs with masses lower than other Seyfert SMBHs. NLS1s are a subclass of broadline Seyfert galaxies. These are AGN detectable through both the active core and the radiation from the host galaxy. They are characterized by extreme multi-wavelength properties, including narrow Balmer lines, strong Fe II emission, and super-soft X-ray emission with extreme variability. NLS1s are thought to accrete at close to Eddington and even super-Eddington rates in some cases \citep{1995MNRAS.277L...5P, 2000MNRAS.314L..17M, 2006AJ....132..531K, 2016MNRAS.455..691J, 2017MNRAS.468.3663J}. In particular, the bolometric luminosity inferred for what is believed to be the most super-Eddington NLS1, RX J0134.2-4258, is firmly in the super-Eddington regime, $L_{\rm bol} \sim (6-16) L_{\rm Edd}$ \citep{2022MNRAS.512.5642J,2023MNRAS.518.6065J}. Furthermore, NLS1s feature steep X-ray spectra that are similar to high-redshift quasars with super-Eddington luminosities. Indeed, NLS1s could be the early-stage of AGN evolution, specifically the low-redshift, low-luminosity versions  of quasars that are actively undergoing super-Eddington accretion.

The centers of many galaxy clusters are observed to be hot, despite their central cooling timescales being much shorter than their lifetimes. BH feedback via jets could be the source of heating that is responsible for avoiding catastrophic cooling (see review by \citealt{2012AdAst2012E...6G}). Centered on the galaxy NGC 1275 is the Perseus cluster, the brightest known X-ray cluster. It is believed that jets inflate bubbles that displace the intracluster medium (ICM) to the North and South of the central BH. These jets, powered by the accretion and spin of the BH, may be the energy source for the observed luminosity \citep{Sanders_2007}, by e.g. driving the turbulence that heats the gas \citep{2014Natur.515...85Z}. 

The galaxy cluster MS 0735 contains an AGN with outflows that extend far beyond the size of the host galaxy, plowing through the ICM and forming large cavities. \cite{2009ApJ...698..594M} argue that because the energy of the outflows is comparable to the rest mass energy of the SMBH, the outflows may be powered in part by the rotational energy of the BH.
For example, the rotational energy of a maximally-spinning $10^9$ solar mass BH is over $10^{62}$ erg. This is large compared to the thermal energy of the surrounding X-ray atmosphere, enough to quench a cooling flow for several Gyrs. 
Accretion power alone is likely not enough to power the jets that heat the ICM, and therefore, they must be powered in part by BH spin energy.  It is now generally accepted that spin power is coupled to the AGN feedback loop \citep{2009ApJ...698..594M}.

It is widely thought that the rotational energy of the BH powers relativistic jets with the help of large-scale magnetic fields through the Blandford-Znajek \citep[BZ,][]{1977BZ, 1969Penrose, 2014PhRvD..89b4041L} process. Naturally, the jet power depends on the value of BH spin, scaling approximately as $P_{\rm jet} \sim \Phi^2 a^2$ \citep{1982MNRAS.198..345M, 1990AN....311..122T,  2012MNRAS.419L..69N}, where $\Phi$ is the magnetic flux on the BH. This scaling is consistent with observations of blazars \citep{2021ApJ...913...93C} and X-ray binaries (\citealt{2012MNRAS.419L..69N,2013ApJ...762..104S,2014SSRv..183..295M}; but see \citealt{2010MNRAS.406.1425F}). 

Radio loudness, which is the total radio luminosity (core plus extended emission) normalized by the B-band nuclear luminosity, shows an apparent spread by factor of $10^3$ between the radio-loud and radio-quiet AGN \citep{2007ApJ_radioloud}. Differences in the accretion rate alone would not be sufficient to explain the dichotomy, because at the same optical luminosity, which is presumably the tracer of accretion rate, the radio-loudness appears to exhibit the dichotomy. One possibility to explain this dichotomy could be that the radio-loud AGN have much more rapidly-spinning BHs than radio-quiet AGN \citep{2007ApJ_radioloud,2010ApJ...711...50T,2011MNRAS.414.1937M,2011ApJ...735...50W}. In this so-called ``spin paradigm'' the value of BH spin controls the jet power \citep{1995ApJ...438...62W, 1999ASPC..160..265B}, as opposed to the ``accretion paradigm'' \citep{1992AIPC..254....3B} where BH accretion rate determines the spectral state of the accretion flow and the jet power. However, in order to achieve the full range of $10^3$ in radio luminosity requires a range of a factor of $10^{1.5}\simeq 30$ in the spin, implying that radio-quiet AGN would have to have a spin $\sim 0.03$, which is increasingly difficult to achieve unless the jet power dependence on spin can be steeper than $a^2$ \citep{2010ApJ...711...50T}. 
 
A BH spin will evolve due to the torques acting on the BH, and over time, the spin will evolve toward an equilibrium value, \aeq. BH spin evolution has been studied for various disk configurations \citep{1970Natur.226...64B, thorne_disk-accretion_1974, 1998ApJ...504..419P, 2004ApJ...602..312G, 2022MNRAS.511.3795N}. Early 2D simulations by \citet{2004ApJ...602..312G} found an equilibrium spin of $\aeq\approx 0.93$, lower than a disk with no large-scale magnetic field. \cite{2012JPhCS.372a2040T,2015ASSL..414...45T,2022MNRAS.511.3795N} found a much lower value of \aeq{} than \citet{2004ApJ...602..312G}, and the effect is due to the presence of strong magnetic fields.

Thanks to recent advances in GRMHD global simulations of accretion disks, it is becoming clear that in the presence of a large amount large-scale vertical magnetic flux, accretion disks naturally end up in a magnetically arrested disk (MAD) state \citep{MAD_tchekho_2011}. Indeed, the large-scale magnetic field threading the disk tends to be dragged inward by the accretion flow. Eventually, the magnetic flux saturates within the inner regions of the disk. This saturated state, the MAD state, is a robust result of GRMHD simulations of accretion disks. Furthermore, it can also occur when the initial magnetic field strength is weak \citep{2021A&A...647A.192J} or the magnetic flux is small \citep{2019MNRAS.490.4811C}.

Evidence is building that some AGN contain MADs with highly efficient jets \citep{2014Natur.510..126Z,2015MNRAS.451..927Z,2015MNRAS.449..316N,2015ASSL..414...45T,2021ApJ...910L..13E, 2021ApJ...913...93C,2022ApJ...930L..16E}. GRMHD models of MADs with high BH spin produce strong jets \citep{MAD_tchekho_2011,2012MNRAS.423.3083M}. Thus, modeling MAD accretion systems could allow us to study the effect the powerful jets have on the BH spin and, hence, jet power. In particular, maximizing jet power, as in MADs, extracts the maximum possible rotational energy from the BH. 

In this paper we use the simulations of \citet{MAD_tchekho_2011,2012JPhCS.372a2040T} to investigate the dynamical differences between the MAD and the geometrically-thin Novikov-Thorne disk \citep{1973blho.conf..343N}, including the impact of jets and their underlying magnetic fields on the evolution of BH spin in time. Whereas the main application of this work is to highly super-Eddington MAD flows, we do not explicitly include radiation transport to model highly super-Eddington accretion disks. In such flows the radiation and gas cannot move relative to each other: effectively, they behave as a single fluid with a polytropic index $\gamma = 4/3$ characteristic of radiation-dominated ideal gas. Thus, we adopt an ideal gas equation of state with a polytropic index of $\Gamma=4/3$. Future work we will include on-the-fly radiation transport to explicitly account for the effect of radiation feedback on the gas (Lowell et al. 2024).
In Section~\ref{sec:style} we give the equations of BH spin evolution and describe the numerical set-up for MAD simulations, where magnetic flux on the BH and jet power are maximized. In Section~\ref{sec:results}, we present our results and derive the semi-analytic model to reveal the physics behind the spin evolution in MADs. In Section~\ref{sec:discussion}, we discuss our results and conclude.

\section{Equations and Method} \label{sec:style}

\subsection{Spin evolution}
\label{sec:maths}

BHs are defined by three parameters: electric charge $Q$, angular momentum $J$, and mass $M$. We assume that BHs are formed from charge-neutral matter, so we work in the Kerr vacuum metric with two parameters, $J$ and $M$.

We define dimensionless BH spin as 
\begin{equation}
-1\le a \equiv \frac{J}{J_{\rm max}} = \frac{J}{M^2} \le 1, 
\label{eq:adef}
\end{equation}
where $J$ is the BH angular momentum, and $M$ is the BH mass. Here and below, we use geometrical units such that $c = G = 1$, and assume that the spin vector is parallel or anti-parallel to the angular momentum vector of the accreting gas.

The change in BH mass-energy $M$ and angular momentum $J$ are given by

\begin{equation}
    dM = e_{\rm in}dm
    \label{eq:dM}
\end{equation} and 
\begin{equation}
    dJ = l_{\rm in}Mdm 
    \label{eq:dJ}
\end{equation} 
\citep{1970Natur.226...64B}, where $dm$ is the amount of accreted rest mass, and $e_{\rm in}$ and $l_{\rm in}$ are the the specific energy and angular momentum, respectively, of that mass.

Combining equations (\ref{eq:dM}) and (\ref{eq:dJ}) gives an ODE that describes spin evolution as a function of BH mass:

\begin{equation}
    \frac{da}{d\log M} = \frac{l_{\rm in}}{e_{\rm in}} - 2a,
    \label{eq:dadlnM}
\end{equation}
where $\log$ is a natural logarithm. Equations (\ref{eq:dM}), (\ref{eq:dJ}), and (\ref{eq:dadlnM}) can be solved for the dependencies $a(m)$ or $a(t)$.

\citet{2004ApJ...602..312G} defined the dimensionless spin-up parameter,

\begin{equation}
    s = \frac{da}{dm}M = \frac{da}{dt} \frac{M}{\dot m},
    \label{eq:spinup_def}
\end{equation} to describe BH spin evolution due to accretion: $s$ is positive when the BH is spinning up, negative when the BH is spinning down, and zero when the BH has reached the equilibrium spin, $\aeq$. 

Using equation (\ref{eq:spinup_def}) we recast equation (\ref{eq:dadlnM}) to find an expression for $s$ as a function of the specific energy and angular momentum fluxes,

\begin{equation}
    s = l_{\rm in}(M,a) - 2a e_{\rm in}(a).
    \label{eq:spinup_defwfluxes}
\end{equation} Here, the angular momentum is positive when the disk is prograde and negative when the disk is retrograde relative to the BH spin. Thus, the accretion of angular momentum $l_{\rm in}$ will lead to  spin-up (spin-down) for prograde (retrograde) disks. Accreted energy is always positive, so the second term spins down (up) the BH for positive (negative) spin. The terms in equation (\ref{eq:spinup_defwfluxes}) are intuitive in that the accretion of positive accreted angular momentum spins up the BH (increases $a$), but the accreted rest-mass spins down the BH.

\subsection{Standard thin disk}
\label{sec:NTdisk} 

In a ``standard'', razor-thin, radiatively-efficient disk model, or Novikov-Thorne (hereafter NT) model \cite{1973blho.conf..343N}, gas moves in Keplerian orbits and experiences viscous forces that cause the gas to lose angular momentum and migrate inward to smaller radii until it reaches the inner-most stable circular orbit (ISCO), also known as the marginally-stable orbit. Inside of the orbit, of radius $R_{\rm ms}$, the gas plunges into the event horizon. In this case the inner radius of the disk is $R_{\rm in}=R_{\rm ms}$. The equations for $e_{\rm in}$, $l_{\rm in}$, and $R_{\rm ms}$ describing the NT disk spin evolution are taken from \citet{Moderski&SikoraBHev}. Analytic expressions for $e_{\rm in}$ and $l_{\rm in}$ can be found in Appendix \ref{sec:appendix_fluxes}.

Physically, the primary origin of viscosity is thought to be the magnetorotational instability \citep[MRI,][]{1991ApJ...376..214B, RevModPhys.70.1}, driven by the magnetic field in the disk. The NT disk model describes the high/soft, or thermal, spectral state of accretion that occurs when the luminosity is $1-10 \%$ of $L_{\rm Edd}$. The model neglects the large-scale poloidal magnetic field: relativistic jets do not form and do not play any role in the spin evolution. The BH spin changes only due to mass accretion onto the BH. This can only increase the angular momentum and energy of the BH, resulting in the spin-up to the maximum spin, $\aeq = 1$. In this work, we ignore the accretion of low angular momentum photons in the NT disk model: including them would limit the BH spin-up to $\aeq = 0.998$ \citep{thorne_disk-accretion_1974}.

\subsection{Simulations of Magnetically Arrested Disks}
\label{sec:simulations}

Unlike for the NT disk, there is no analytic model to describe MAD spin evolution. Therefore, numerical simulations of MADs must be used. We use non-radiative 3D GRMHD simulations run with the code HARM \citep{2003ApJ...589..444G,2006ApJ...641..626N} of BH accretion in the MAD regime with BH spins $-0.9$, $-0.5$, $-0.2$, $0.0$, $0.1$, $0.2$, $0.5$, $0.9$, and $0.99$ \citep{MAD_tchekho_2011,2012JPhCS.372a2040T}. In order to make inferences for super-Eddington accretion, we adopt an ideal gas law equation of state with a polytropic index of $\Gamma = 4/3$, characteristic of a radiation-dominated gas. The initial conditions are such that an accretion disk will form with scale height $h/r \approx 0.3$, and enough magnetic flux $\Phi_{\rm BH}$ will accumulate on the BH to obstruct the accreting gas and lead to a MAD. The simulation parameters are given in Table \ref{tab:sim_dets}. Because BH mass and spin evolve over cosmological time, much longer than the duration of any GRMHD simulation, we ``stitch'' together simulations for different BH spin values to represent the BH spin evolution over cosmological time. Our simulations reach the quasi-steady MAD state by $t \sim 10,000 r_{\rm g} /c$. From here we time-average quantities to the end of each simulation, as shown in Table \ref{tab:sim_dets}, to reduce the effect of fluctuations due to magnetic flux eruptions. The simulations reach inflow equilibrium out to $(20-50) r_{\rm g}$ within the time-average, where the distance depends on the simulation.
 
 \begin{table*}
     \centering
     \caption{Simulation details. $l_{\rm MAD}$ and $e_{\rm MAD}$ are the $t,\theta,\phi$ -averaged angular momentum and energy fluxes, respectively, on the horizon.}
     \begin{tabular}{c|c|c|c|c|c|c|c}
     \hline
     $a$ & Resolution & $\Delta_\phi$ &$r_{\rm in}/r_g$ & $r_{\rm max}/r_g$ & $t_{\rm avg}$ & $l_{\rm MAD}$ & $e_{\rm MAD}$\\
        &   ($N_r \times N_\theta \times N_\phi$) &  & & &  $(r_g/c)$ & & \\
    \hline
         -0.9 &  $288\times 128\times 64$ & $2\pi$ & 15 & 34 & ($10000; 20100$) & 4.77 & 0.668 \\
         -0.5 &  $288\times 128\times 32$ & $\pi$ & 15 & 36.21 & ($10000; 16350$) & 4.50 & 0.830 \\
         -0.2 &  $288\times 128\times 32$ & $\pi$ & 15 & 35.64 & ($10000; 15200$)  & 3.23 &  0.931 \\
         0.0  &  $288\times 128\times 32$ & $\pi$ & 15 & 35 & ($10000; 18550$)   & 1.03 & 0.954 \\
         0.1  &  $288\times 128\times 32$ & $\pi$ & 15 & 35 & ($10000; 18725$)   & -0.31 & 0.946 \\ 
         0.2  &  $288\times 128\times 32$ & $\pi$ & 15 & 35 & ($10000; 13400$)   & -1.20 & 0.907   \\
         0.5  &  $288\times 128\times 32$ & $\pi$ & 15 & 34.475 & ($10000; 13050$)   & -3.92 & 0.699 \\
         0.9  &  $288\times 128\times 64$ & $2\pi$ & 15 & 34.1 & ($10000; 19900$)    & -7.48 & -0.006 \\
         0.99 &  $288\times 128\times 64$ & $2\pi$ & 15 & 34   & ($10000; 14650$)  & -8.47 & -0.541 \\
    \hline
     \end{tabular}
     \label{tab:sim_dets}
\end{table*}

We calculate the specific angular momentum and energy fluxes from the simulations using the stress-energy tensor, defined as

\begin{equation}
    T^{\mu \nu} = T^{\mu \nu}_{\rm HD} + T^{\mu \nu}_{\rm EM}.
    \label{eq:Tmunu}
\end{equation} The hydrodynamic component is given by 

\begin{equation}
    T^{\mu \nu}_{\rm HD} = (\rho + u + p)u^\mu u^\nu + p g^{\mu \nu}
    \label{eq:Tmunu_hydro}
\end{equation} and is used to calculate specific fluxes $e_{\rm in}$ and $l_{\rm in}$ on the BH from the disk. Here, $\rho$ is the gas density, $u$ is the specific internal energy and $p$ is thermal pressure of the gas, and $u^\mu$ is the gas contravariant 4-velocity. Electromagnetic (EM) fluxes are calculated using the electromagnetic component of the stress-energy tensor, given by

\begin{equation}
    T^{\mu \nu}_{\rm EM} = b^2 u^\mu u^\nu + \frac{1}{2} b^2 g^{\mu \nu} - b^\mu b^\nu,
    \label{Tmunu_EM}
\end{equation}
where $p_{\rm mag} = b^2/2 = b^\mu b_\mu/2$ is the magnetic pressure and $b^\mu$ is the comoving magnetic field 4-vector.
The mass, specific energy, and specific angular momentum fluxes are calculated as 

\begin{equation}
    \dot{m}(r) = \iint \rho u^r dA_{\theta \phi}, 
    \label{eq:fM}
\end{equation}

\begin{equation}
    e(r) = \frac{\dot{E}(r)}{\dot{m}(r)} = \frac{1}{\dot{m}(r)} \int_\theta \int_\phi T^r_t dA_{\theta \phi}
    \label{eq:fEoverfM},
\end{equation} and

\begin{equation}
    l(r) = \frac{1}{M} \frac{\dot{L}(r)}{\dot{m}(r)} = \frac{1}{M} \frac{1}{\dot{m}(r)} \int_\theta \int_\phi T^r_\phi dA_{\theta \phi},
    \label{eq:fLoverfM}
  \end{equation}
where $dA_{\theta \phi}$ is the area element in the $\theta-\phi$ plane.

The torques on the BH are due to accreted fluxes, so in our simulations we calculate the fluxes at the BH event horizon, $r = r_{\rm H} = r_g(1+\sqrt{1-a^2})$, where $r_g = GM/c^2$ is BH gravitational radius. The total, electromagnetic, and hydrodynamic specific angular momentum fluxes are calculated at $r_{\rm H}$ as $l_{\rm in}=l(r=r_{\rm H})$, as is the electromagnetic specific energy flux. 

However, when calculating hydrodynamic fluxes we must account for numerical floors. Realistically, there is no mass in the jet, but numerical MHD schemes do not work in a vacuum. To handle this, when mass or internal energy densities are very small, below a set floor value, internal energy or mass is added to the grid near the BH horizon to keep the numerical scheme stable. When calculating the mass, hydrodynamic energy and angular momentum fluxes, we eliminate contributions from the highly magnetized regions affected by the floors using a magnetization cutoff condition. This method gives the physical values at the BH horizon. Our treatment of the floors on the hydrodynamic fluxes is described further in Appendix~\ref{sec:appendix_floors}.

\section{Results}
\label{sec:results}

\begin{figure}
    \centering
    \includegraphics[width=\columnwidth]{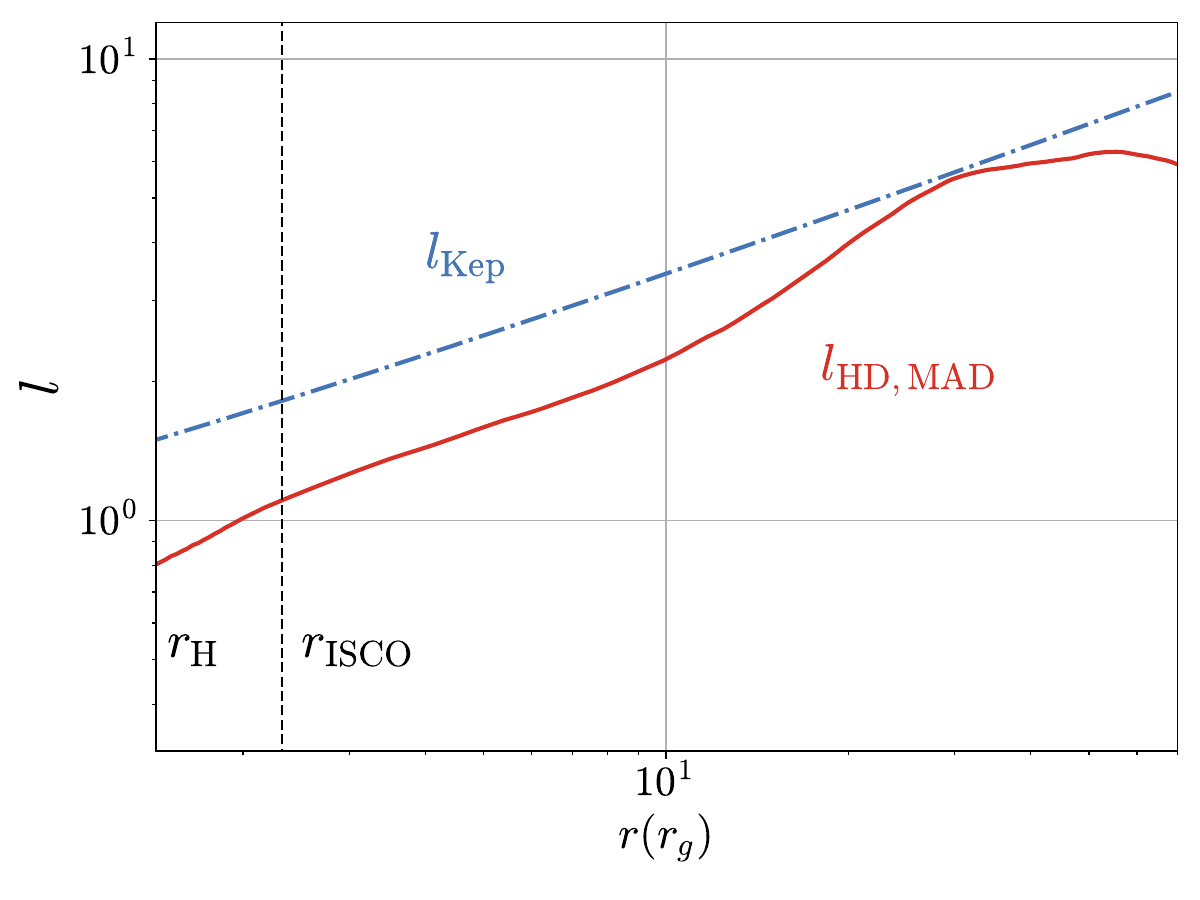}
    \caption{
    Non-relativistic Keplerian specific angular momentum (blue; $l_{\rm Kep}=r^{1/2}$) and MAD (red) normalized angular momentum flux vs. radius for $a=0.9$. The BH horizon radius $r_{\rm H}$ is the lower x-axis limit, and the innermost stable circular orbit radius $r_{\rm ISCO}$ is shown by the vertical dashed line. The angular momentum in the hydrodynamic flow in the MAD, $l_{\rm HD, MAD}$, is sub-Keplerian as the gas falls onto the BH, indicating that some mechanism is robbing the disk of angular momentum thereby depriving the BH of the angular momentum supply. }
    \label{fig:Ang_Kep_drop}
\end{figure}

We compared the angular momentum supply at the BH event horizon radius for the thin disk model and the MAD simulations. We found that the hydrodynamic disk in the MAD simulations does not behave in the same way as the NT disk. Figure \ref{fig:Ang_Kep_drop} shows  that the infalling material in the inner accretion disk becomes strongly sub-Keplerian and sub-NT. Strikingly, specific angular momentum flux on the BH, computed using the hydrodynamic component of $T^{r}_{\phi}$ (Eq.~\ref{eq:fLoverfM}), makes up only $37 \%$ of the NT value, a consequence of strong magnetic fields in MADs. This value is strikingly low even for MADs, because usually MAD angular momentum is compared to the Keplerian value, with the typical degree of MAD sub-Keplerian rotation quoted at around $50$\% \citep{2008ApJ...677..317I,2012MNRAS.423.3083M,2022MNRAS.511.2040B}. Hence, the disk has nearly three times less specific angular momentum than the NT expectation to offer the BH and might not be able to significantly contribute to spinning up the BH.

\subsection{BH spin evolution for NT and MAD flows}
\label{sec:results_spinev}

\begin{figure}
    \centering
    \includegraphics[width=1.0\columnwidth, angle=0]{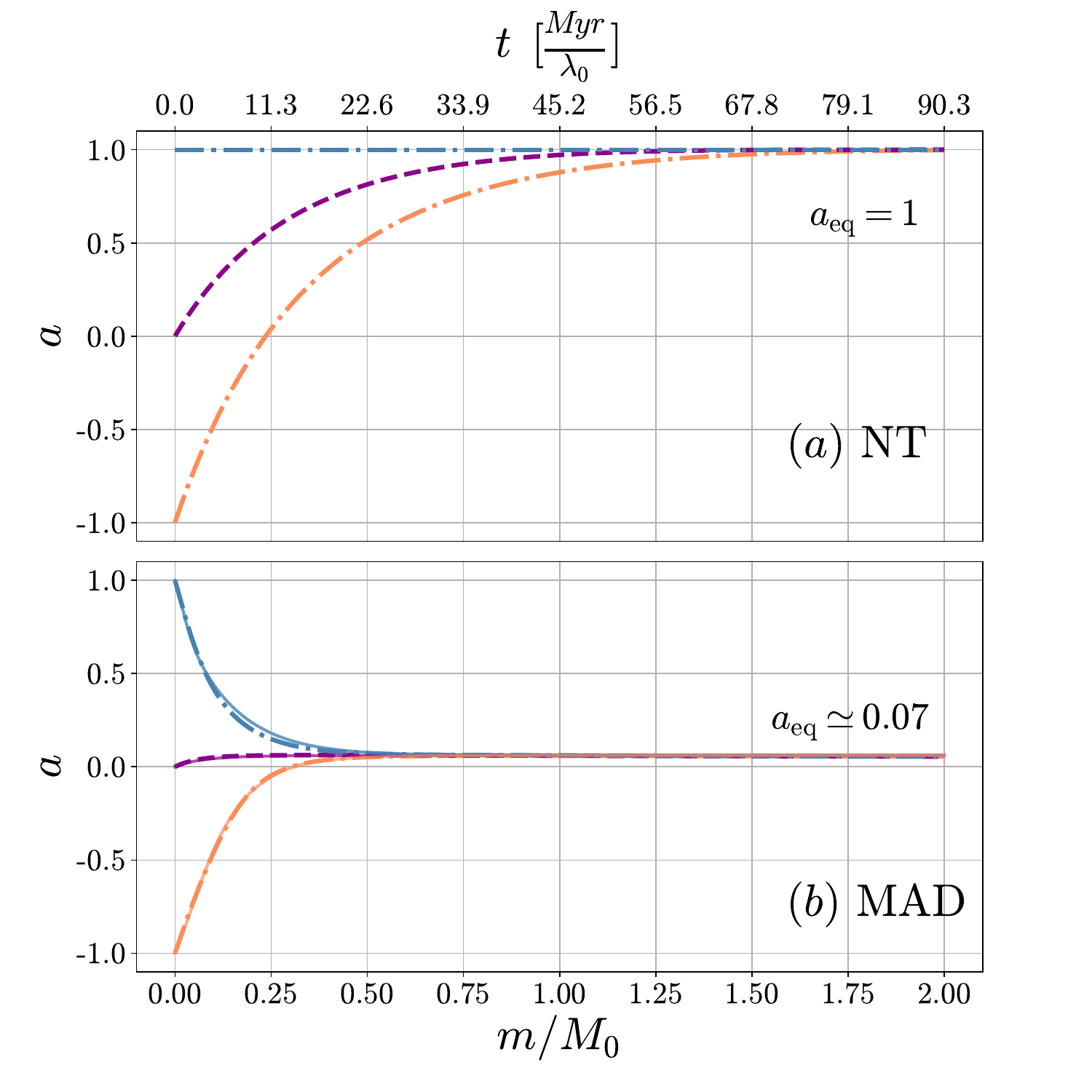}
    \caption{Dimensionless spin $a(m)$ and $a(t)$ for different values of initial spin: $a_0 = -1$, $0$, $1$. The lower $x$-axis gives the mass accreted over the initial BH mass: when $m/M_0=1$, the BH has accreted its entire initial mass. The upper $x$-axis gives spin evolution in Myrs over an initial Eddington rate $\lambda_0$. [panel (a):] The BH with a Novikov-Thorne (NT) disk gains angular momentum and energy, and will spin up to $\aeq=1$. [panel (b):] A MAD will spin down the BH to $\aeq \approx 0.07$. The MAD spin-down timescale is very short (much shorter than NT spin-up timescale): the BH reaches its equilibrium spin after accreting only half of its initial mass. For a fixed $\lambda_0$, this spin-down timescale is independent of initial BH mass. Dashed lines show BH spin evolution based on our GRMHD simulation results (computed using a spline interpolation of simulated data points, $e_{\rm in}(a)$ and $l_{\rm in}(M,a)$, shown in Fig.~\ref{fig:lin_ein}) and solid thin lines show the spin evolution computed using the semi-analytic model described in Sec.~\ref{sec:results_MADmodel}: our model provides a good description of the simulation results.}
    \label{fig:spinev}
\end{figure}

\begin{figure}
    \centering
    \includegraphics[width=1.0\columnwidth]{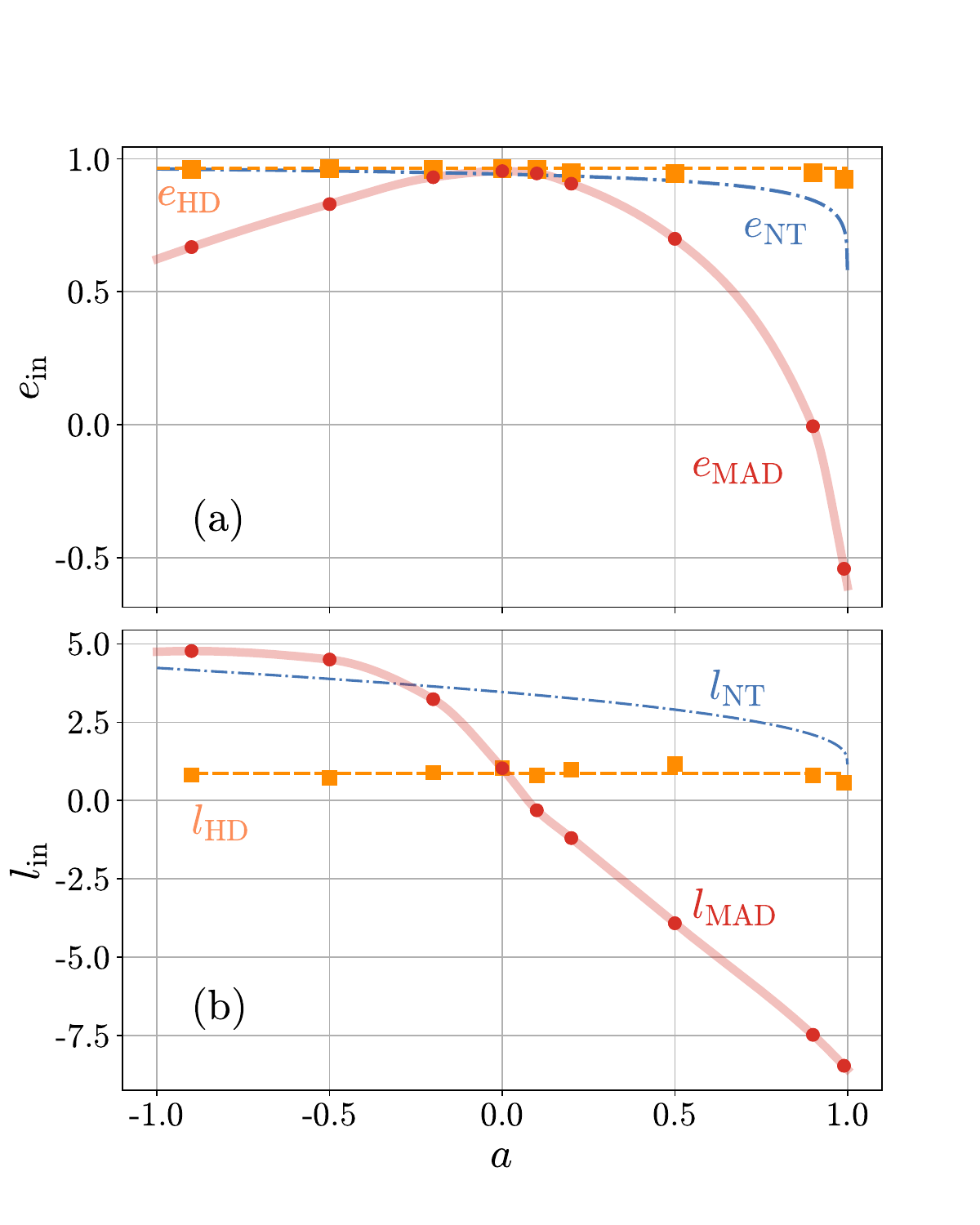}
    \caption{Specific energy and angular momentum fluxes in the Novikov-Thorne and MAD models. [panel (a):] Specific energy flux: The Novikov-Thorne energy flux, $e_{\rm NT}$, on the BH remains positive and decreases rapidly at high prograde spin (blue dash-dotted curve). In a MAD, the hydrodynamic portion, $e_{\rm HD}$, stays close to unity (orange dashed line), although it shows a slight decrease near $a = 1$ (orange squares).     
    Additionally accounting for the EM contribution leads to the total specific MAD energy flux, $e_{\rm MAD}$ (red dots connected by light red line).
    [panel (b):] In the NT model, the specific angular momentum flux, $l_{\rm in, NT}$, decreases with increasing spin (blue dash-dotted line). Surprisingly, in a MAD, $l_{\rm in, HD}$ is much smaller than the NT value (orange squares) and roughly constant at all spins: 
    the MAD specific angular momentum flux is significantly sub-NT. In fact, the hydrodynamic component of the angular momentum flux is essentially constant across all values of spin (orange dashed line). 
    In our semi-analytic model of MAD spin-down, we use the two (constant) fits shown with the dashed orange lines, $l_{\rm HD}$ and $e_{\rm HD}$. The thickness of the red lines following $l_{\rm MAD}$  and $e_{\rm MAD}$ data points does not represent uncertainty. The red lines show the spline interpolation through the simulated data points, and horizontal orange lines in panels (a) and (b) represent the best-fits for $e_{\rm HD}$ and $l_{\rm HD}$, respectively, that we use in our semi-analytic model shown in Fig.~(\ref{fig:spinup_param}).}
    \label{fig:lin_ein}
\end{figure}

To determine BH spin evolution, we solved the coupled ODEs, equations (\ref{eq:dM}), (\ref{eq:dJ}), and (\ref{eq:dadlnM}), for BH spin $a(m)$ and mass $M(m)$, for both the standard NT disk and MAD flows. Figure~\ref{fig:spinev} shows the BH spin evolution $a(m)$ for the NT (Fig.~\ref{fig:spinev}a) and MAD (Fig.~\ref{fig:spinev}b) cases. The NT evolution is calculated using the equations described in Section~\ref{sec:maths}. From each MAD simulation, we calculate the fluxes at the horizon, $e_{\rm in}$ and $l_{\rm in}$. We then use spline interpolation to evaluate the simulated data points over the full range of spins, resulting in the interpolation functions, $e_{\rm in}(a)$ and $l_{\rm in}(M,a)$, that we plug into the ODEs (see Fig.~\ref{fig:lin_ein}). We then solve the ODE equations to model the spin evolution.

For each disk type, we consider three values for the initial spin: $a_0 = -1$,  $a_0 = 0$, and  $a_0 = 1$. For any initial spin $a_0$, the NT disk in Figure~\ref{fig:spinev}(a) will spin up the BH to equilibrium at the maximum $\aeq = 1$ before the BH has accreted twice its initial mass, $m/M_0 = 2$. The infalling gas supplies angular momentum and energy to the BH, and in the NT model there is nothing to remove angular momentum or energy from the BH. Thus, $a$ can only increase. For the MAD case shown in Figure~\ref{fig:spinev}(b), the BH spins down rapidly to the low value of $a_{\rm eq,MAD} \approx 0.07$, and in doing so it has only accreted roughly $50 \%$ of its initial mass. 

We translate this timescale into physical units at the upper $x$-axis in Fig.~\ref{fig:spinev}. The Eddington accretion rate is given by

\begin{equation}
    \dot{M}_{\rm Edd} = \frac{4 \pi G m_p M}{\epsilon \sigma_{\rm T} c},
    \label{eq:Mdot_Edd}
\end{equation} where $\sigma_{\rm T}$ is the Thomson cross-section, $m_p$ is the proton mass, and $\epsilon$ is the radiative efficiency. We assume $\epsilon = 0.1$. We show the cosmological spin-down timescale in units of Myrs over a fixed Eddington ratio,

\begin{equation}
    \lambda_{0} = \frac{\dot{m}}{\dot{M}_{\rm Edd}}.
    \label{eq:Edd_ratio}
\end{equation} For a BH accreting at the Eddington rate ($\lambda_0 = 1$), the MAD BH will spin down from $a=1$ to $a=0.2$ in approximately $9$ Myrs, by which time it will consume about $20$\% of its own mass.

In order to explain how the equilibrium spin in the MAD case, $a_{\rm eq, MAD}$, can become so small, as well as why the spin-down is so much faster compared to the NT case, we need to understand the physics of the torques acting on the BH. First, we know that MADs launch powerful electromagnetic jets, and from the BZ model we know that they are powered by BH spin. Additionally, the accretion disk will exert a torque on the BH. Each in part carry their own angular momentum and energy. We now identify the influence of each of these factors.

\subsection{BH power output}
\label{sec:results_jetefficiency}

We showed that in the NT disk, the BH can only spin up with time. The BH eats energy and angular momentum, and there are no large-scale magnetic fields to efficiently carry away the angular momentum. In a MAD, energy and angular momentum do not travel just into the horizon, but also outward in the form of jets and winds. They derive their power in part from the EM BZ power of the rapidly spinning BH. We define the EM power efficiency as the ratio of the time-averaged EM energy flux leaving the BH to the time-averaged rest mass flux entering the BH: 
\begin{equation}
    \eta_{\rm EM} = \frac{\langle P_{\rm EM} \rangle}{ \langle \dot m \rangle c^2} .
    \label{eq:eta_EM}
\end{equation}

\begin{figure}
    \centering
    \includegraphics[width=1.0\columnwidth]{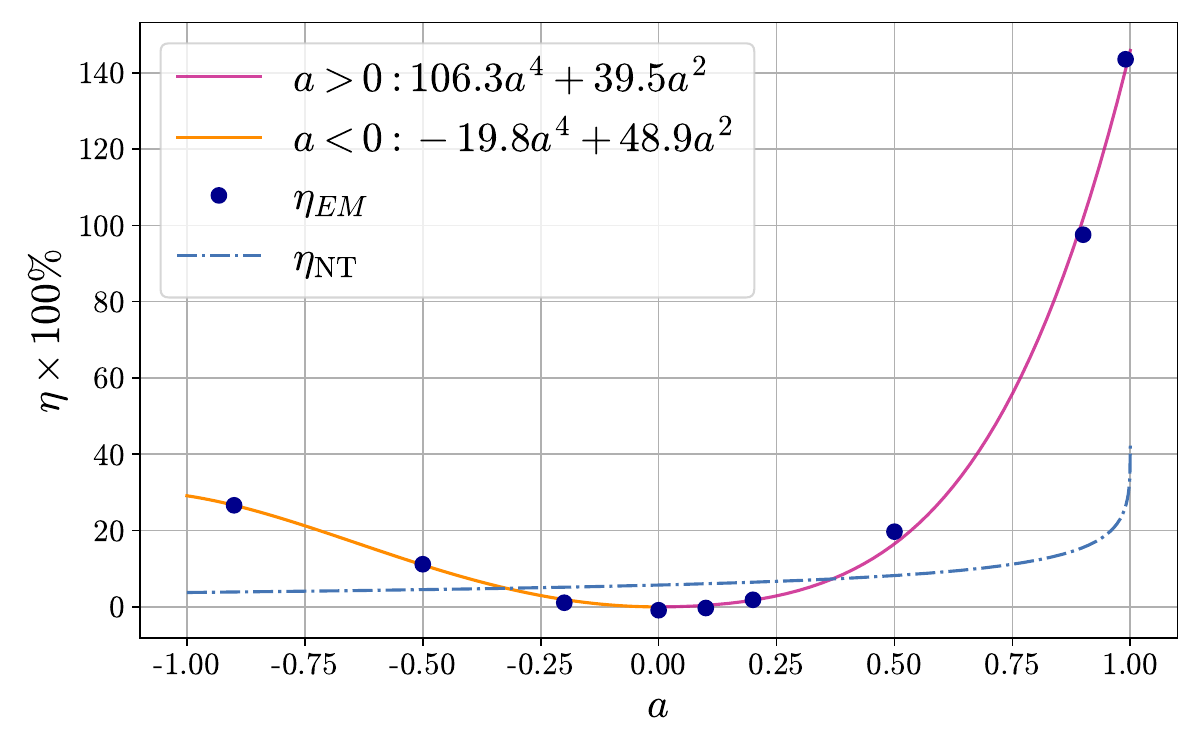}
    \caption{Dimensionless efficiency $\eta$ as a function of BH spin $a$ for different models. The radiative efficiency $\eta_{\rm NT}$ for the Novikov-Thorne disk is shown in the blue dash-dotted line, and the electromagnetic outflow efficiency $\eta_{\rm EM} = P_{\rm EM}/\dot{m}c^2$ at the event horizon is shown with dark blue dots. We provide separate fits to $\eta_{\rm EM}(a)$ dependence for positive (pink) and negative (orange) spins (see eq.~\ref{etafit}). A rapidly-spinning BH fed by a MAD can achieve EM energy outflow efficiencies $> 100 \%$. These fits form the inputs to our semi-analytic MAD spin-down model.}
    \label{fig:eta}
\end{figure}

We plot the EM efficiency $\eta_{\rm EM}$ as a function of spin in Figure \ref{fig:eta}. The electromagnetic energy flux on the BH is used to calculate the jet efficiency for each simulation and is shown as black points. To produce the pink and orange curves, we use the least squares fits in the form $C_1a^4 + C_2 a^2$ for each of the negative and positive values of $a$. This results in the fit for $\eta_{\rm EM}$:
\begin{equation}
\eta_{\rm EM} \times 100 = \left\{
\begin{array}{ll}
      -19.8 a^4 + 48.9 a^2, & a \leq 0, \\
      106.3 a^4 + 39.5 a^2, & a \geq 0. \\
\end{array} 
\label{etafit}
\right.
\end{equation}

For a NT disk, $\eta_{\rm NT}$ never exceeds $45 \%$. In the work of \cite{2004ApJ...602..312G}, the torque on the central BH was provided by a disk model that reached a maximum jet efficiency of $16 \%$. MADs are known to achieve enormous efficiencies of $150 \%$ \citep{MAD_tchekho_2011} and even higher \citep{2012MNRAS.423.3083M}. An efficiency of $100 \%$ means that all accretion power is processed into jet power. Thus an efficiency exceeding $100 \%$ implies that the jets are feeding on the angular momentum of the BH.

For the BH to launch jets, it needs to hold magnetic flux. Jet power is directly proportional to magnetic flux, $\Phi_{\rm BH}$. BH generates jet power via the BZ mechanism at the efficiency \citep{1977BZ,2015ASSL..414...45T},
\begin{equation}
    \eta_{\rm BZ} =  \frac{\langle P_{\rm BZ} \rangle}{ \langle \dot m \rangle c^2} = \frac{\kappa}{4 \pi c} \left(\frac{\Omega_{\rm H} r_g}{c}\right)^2
    \langle \phi^2_{\rm BH} \rangle f(\Omega_{\rm H}),
    \label{eq:eta_BZ}
\end{equation} where $\kappa$ is a constant that depends on the magnetic field geometry (0.044 for a parabolic geometry), $\Omega_{\rm H} = ac/(2r_{\rm H})$ is the angular frequency of the BH horizon, $r_{\rm H}$ is the radius of the BH horizon, and $\Phi_{\rm BH}$ is the magnetic flux on the BH. The approximation, $f(\Omega_{\rm H}) = 1$, can be used for $a \leq 0.95$, but for rapidly-spinning BHs with $a > 0.95$, the following approximation maintains accuracy, $f(\Omega_{\rm H}) \approx 1 + 1.38(\Omega_{\rm H} r_{\rm g}/c)^2 - 9.2(\Omega_{\rm H} r_{\rm g}/c)^4$ \citep{2015ASSL..414...45T}. We use $\eta_{\rm BZ}$ to guide our choice of fit for $\eta_{\rm EM}$ in Figure \ref{fig:eta} and equation \eqref{etafit}.

\subsection{Disk and jet contributions to spin-down}
\label{sec:results_specificfluxes}

Both the disk and the jets exert a torque on the BH. To understand why the MAD spins down to such a low value of spin and why it happens so quickly, we now decompose the total torque into its EM and hydrodynamic constituents, which we compute via the corresponding fluxes at the event horizon.
We distinguish the disk (from the jets) using a cutoff condition on the magnetization, $\sigma\equiv 2p_{\rm mag}/\rho c^2 < 30$; in the simulations, we set the density floor, which limits the magnetization in the polar regions to $\max\sigma=50$, see Appendix \ref{sec:appendix_floors} for more details. Here, $p_{\rm mag}$ is the magnetic pressure. Using the above condition, we find that, unsurprisingly, most of the electromagnetic (EM) flux escapes through the jets, and most of the hydrodynamic flux enters through the disk. To keep things simple, from now on, we will assign the torque by the jets to be the entire electromagnetic component and the torque by the disk to be the entire hydrodynamic component. We note that despite this naming convention, some ``jet'' magnetic field lines may connect to the BH through the disk. 

We plot specific angular momentum and energy fluxes vs.\ spin in Figure~\ref{fig:lin_ein}. Figure~\ref{fig:lin_ein}(a) shows that the NT disk (blue dashed line) supplies the BH the most angular momentum flux and energy flux at $a=-1$ and the least at $a=1$, and always supplies positive energy and angular momentum. The total MAD energy flux computed from our simulations, $e_{\rm MAD}$, is shown in red, where each data point corresponds to a simulation. We find that its hydrodynamic constituent remains roughly constant, $e_{\rm in, HD} \approx 0.97$, 
for all but the highest positive values of $a$. However, at such high values of $a$, it is the EM component that dominates $e_{\rm MAD}$. Figure~\ref{fig:lin_ein}(b) reveals that the hydrodynamic specific angular momentum (orange) supplied by the MAD is significantly less than that supplied by the NT disk. The difference in angular momentum supply might be related to the strong magnetic fields efficiently transporting the disk angular momentum outwards, either within the disk or in the form of large-scale outflows (e.g., \citealt{blandford_hydromagnetic_1982}). 
Furthermore, Fig.~\ref{fig:Ang_Kep_drop} shows that the disk rotation becomes significantly sub-Keplerian, which can starve the BH of the hydrodynamic angular momentum. This sub-Keplerian rotation can be related to the vertical thickness of the disk but it is probably also related to the large scale field \citep{scepi_magnetic_2023}.

It is striking that $l_{\rm in, HD}$ remains roughly constant at $\approx 0.86$ for all values of $a$. However, it can be understood based on the argument described above. The dominant contributions to $l_{\rm HD}$ are the strong field and the sub-Keplerian motions. Both effects will reduce the importance of the ISCO, which is defined for Keplerian motions, leading to $l_{\rm HD}$ being independent of spin and only dependent on the disk dynamics.

Interestingly, the MAD spin-down cannot be reproduced by simply taking the torque from the NT disk and adding the spin-down contribution from the jets. In fact, we cannot assume that the two disks supply the same angular momentum and energy to the BH. In fact, the MAD hydrodynamic torque is quite different than that of the NT disk. This is one reason for why the MAD spins the BHs down to $\aeq \approx 0.07$, and why it happens so quickly: at every stage of its evolution, a MAD supplies very little angular momentum to the BH. 

\begin{figure}
    \centering
    \includegraphics[width=\columnwidth]{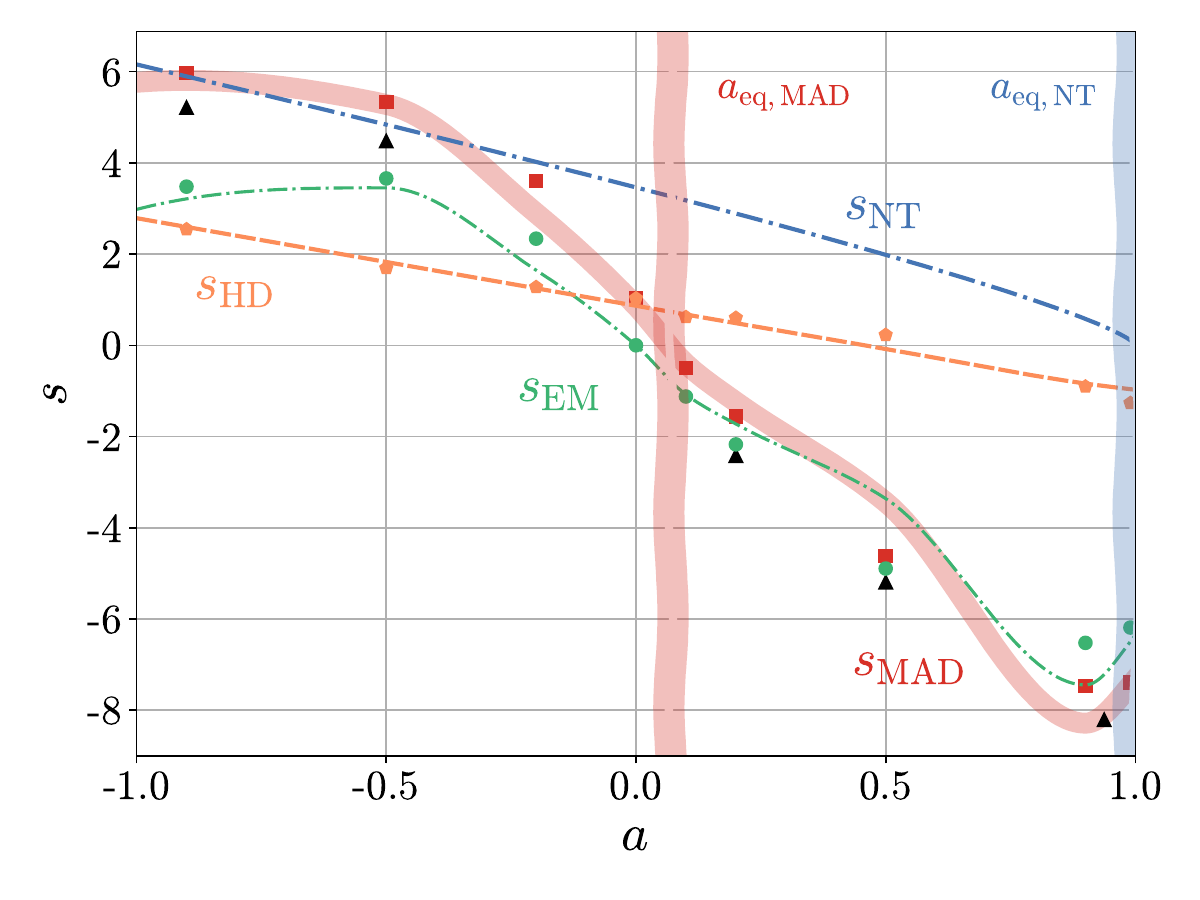}
    \caption{Dimensionless spin-up parameter $s$ as a function of BH spin $a$ for MAD and NT disk models. We show the NT model in blue. MAD hydrodynamic disk is shown in orange, and our model for the disk is shown by the dashed orange line. In a MAD, the disk component follows a similar trend as the NT disk, but its torque on the BH is much smaller except at the highest spin values. The EM component is shown in green, with our jet model shown by the green dashed line. The total MAD data is shown by red points, and our total MAD spin-down model is shown by the thick, red curve. The values of equilibrium spin $\aeq$ for MAD and NT disks are shown as vertical lines at $a=0.1$ and $a=1$, respectively: these are where the spin-up curves vanish, $s=0$. We note that the thickness of the vertical lines does not correspond to uncertainty. Our semi-analytic $s_{\rm MAD}$ model agrees well with the simulation data. Differences for positive spin are primarily caused by an imperfect fit for $\eta_{\rm EM}$ (as seen in Fig.~\ref{fig:eta}), affecting $s_{\rm EM}$. Our semi-analytic model for MAD torques on the BH shows how and why MADs spins down their BHs to a very low value of $a_{\rm eq,MAD}\approx 0.07$. MADs are significantly modified by the large-scale, dynamically-important magnetic fields. Black triangles are higher resolution simulations with $\Gamma=13/9$, demonstrating that higher $\Gamma$ leads to stronger spin-down.
    }
    \label{fig:spinup_param}
\end{figure}

We illustrate the MAD spin-down in Figure \ref{fig:spinup_param} through the dimensionless spin-up parameter \citep{2004ApJ...602..312G}, which is defined in equation (\ref{eq:spinup_def}). When $s=0$ the BH spin is no longer evolving and has reached equilibrium. The blue dashed line shows the spin-up parameter for the thin Novikov-Thorne disk, where the blue vertical band shows where $s_{\rm NT}=0$ at $a=1$, just as we showed in Figure \ref{fig:spinev}. The red curve shows the total spin-up from the MAD simulations and was calculated using the form of equation (\ref{eq:spinup_defwfluxes}) with equations (\ref{eq:fEoverfM}) and (\ref{eq:fLoverfM}). The red vertical band shows $s_{\rm MAD} = 0$ at $a \approx 0.1$, just as we showed $a_{\rm eq, MAD} \approx 0.1$ in Section \ref{sec:results_spinev}. The spin-up parameter remains positive for a NT disk, since accretion supplies angular momentum and energy, increasing the BH spin. For a MAD with $s < 0$, the BH loses energy to relativistic jets and its spin decreases back down to the equilibrium value.

We note that several simulations, e.g. those with $a=0.2$ and $a=0.5$, have a relatively short duration, which limits the time interval over which we averaged them over to measure the spin-up parameter (Table~\ref{tab:sim_dets}). However, the simulations that bracket the equilibrium spin, those for $a=0.0$ and $a=0.1$, extend out to later times. To compute the equilibrium spin numerically, we use linear interpolation between the simulated values of $s$ for these two longer simulations.

Orange points in Figure~\ref{fig:spinup_param} show the hydrodynamic component of the spin-up parameter, $s$. Interestingly, $s_{\rm HD}$ becomes negative at $a\gtrsim0.5$. Although the disk supplies the BH with angular momentum, at high spin the magnetic field removes much of the angular momentum from the disk before it reaches the BH. From Eq.~\eqref{eq:spinup_defwfluxes}, we see that the value of $s$ comes from the balance between the energy and angular momentum fluxes into the BH. Because in the definition of $a$ (Eq.~\ref{eq:adef}) the angular momentum enters in the numerator ($J$), the supply of positive angular momentum increases the BH spin. In contrast, BH mass ($M$) enters in the denominator, so its increase (through the supply of positive energy) reduces the BH spin. Because the energy supply term in Eq.~\eqref{eq:spinup_defwfluxes} is proportional to BH spin, its effect increases with increasing spin and ends up dominating the angular momentum term at high spin: in other words,  at high spin the disk angular momentum supply to the BH is too low to balance out the energy supply. This causes the hydrodynamic torque on the BH to become negative at high spin values.

\subsection{Model for MAD spin-down}
\label{sec:results_MADmodel}

Here we construct a model that reproduces the MAD simulation data. Jet power must play an important role in BH spin-down because the jets extract BH rotational energy at high spin. \citet{Moderski&SikoraBHev} described the spin evolution, including the effect of jet-like outflows, analytically by taking the BH spin evolution equations (Eqns. \ref{eq:dM}  and \ref{eq:dadlnM}) and including the terms for jet power. We write them as
\begin{equation}
    \frac{da}{dt} = \frac{\dot{m}}{M} \left(l_{\rm in, HD} - 2a e_{\rm in, HD}\right) - \frac{P_{\rm EM}}{M c^2} \left(\frac{1}{k \Omega_{\rm H}} - 2a \right)
    \label{eq:dadt_diskjet}
\end{equation} 
and 
\begin{equation}
    \frac{d \log M}{dt} = \frac{\dot{m}}{M} e_{\rm in, HD} - \frac{P_{\rm EM}}{M c^2},
    \label{eq:dlnMdt_diskjet}
\end{equation} where $ k = \Omega_{\rm F} / \Omega_{\rm H}$, the ratio of the angular frequency of the magnetic field lines to the angular frequency of the BH event horizon, $\Omega_{\rm H}$. The first terms now only describe the disk (hydrodynamic) contributions.

As we did for the NT disk and MAD, we similarly recast equation (\ref{eq:dadt_diskjet}) in the form of equation (\ref{eq:spinup_def}) to now determine how the disk and jets contribute to BH spin-up. The new expression for $s$ is then
\begin{equation}
    s = \left(l_{\rm in, HD} - 2a e_{\rm in, HD} \right) - \eta_{\rm EM} \left(\frac{1}{k \Omega_{\rm H}} - 2a\right).
    \label{eq:spinup_diskjet}
\end{equation} 

\begin{figure}
    \centering
    \includegraphics[width=0.9\columnwidth]{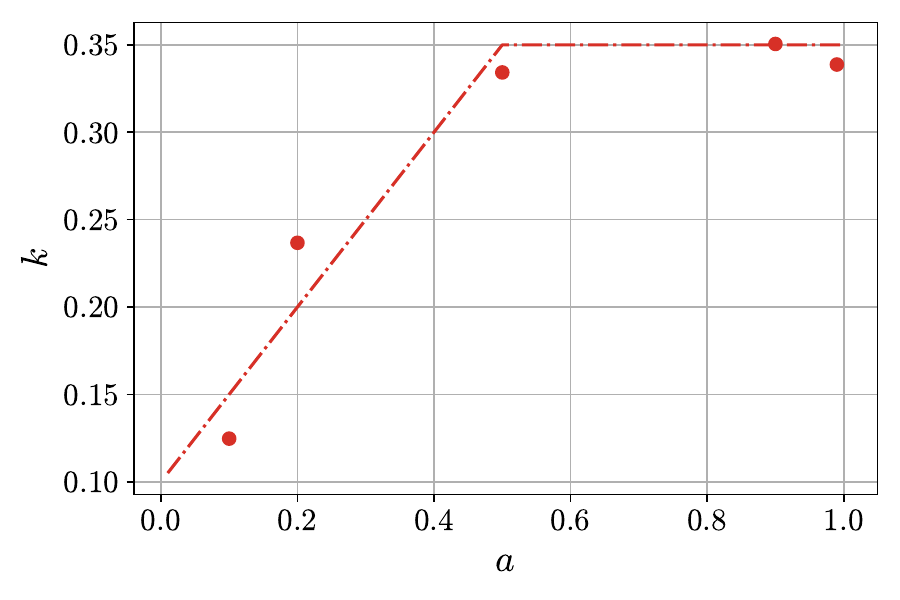}
    \caption{ We plot $k=\Omega_{\rm F} / \Omega_{\rm H}$  and its dependence on spin, which is crucial for converting jet power into angular momentum in our semi-analytic model. The red dashed line shows the fit that we used to construct our semi-analytic model for $a>0$. We did not find a strong dependence on spin for $a<0$ and so used a constant value, $k=0.23$, in our semi-analytic model for $a<0$.}
    \label{fig:kvsa}
\end{figure}

We now use approximate fits for $k$, $\Omega_{\rm H}$, $\eta_{\rm EM}$, $l_{\rm in, HD}$, and $e_{\rm in, HD}$ (see Figs.~\ref{fig:eta}, \ref{fig:lin_ein}, and \ref{fig:kvsa}) to construct our semi-analytic model. Namely, we approximate $l_{\rm in, HD}$ with the constant value, $l_{\rm HD,avg} = 0.86$, and $e_{\rm in, HD}$ with the average value of $0.97$, both shown by the dashed orange lines in Figure~\ref{fig:lin_ein}. We use the fit, $\eta_{\rm EM}$, in the form $\eta \propto C_1 a^4 + C_2 a^2$ as we show in Figure~\ref{fig:eta} and equation~(\ref{eq:eta_BZ}). 

The standard monopole BZ model has $k = 1/2$, which gives the maximum EM efficiency.  We calculate $k$ at the horizon and then average in $t$, $\theta$ and $\phi$. Figure~\ref{fig:kvsa} shows that for our semi-analytic model, we approximate this dependence with the average value, $k = 0.23$, for $a < 0$. For $a > 0$, we find a good fit for $k$,
\begin{equation}   
k = \left\{
\begin{array}{ll}
      0.23, & a < 0, \\
      \min(0.1+0.5a, 0.35), & a > 0. \\
\end{array} 
\right.
\label{eq:kdef}
\end{equation} 
Similar to \citet{2013MNRAS.Penna}, $k$ is smaller than what is predicted by the standard BZ model. 

We could have modeled the angular momentum extraction by the jet EM torques by fitting $l_{\rm EM}$ instead of $k$. However, $k$ is a useful quantity in BH jet theory because it connects the angular momentum extraction efficiency to the better-known and better-calibrated EM energy extraction efficiency, $\eta_{\rm EM}$. Moreover, from the consideration of the standard BZ effect, we expect $k\sim0.25-0.5$, for a wide range of jet geometries ranging from monopolar to parabolic \citep{2010ApJ...711...50T}. Therefore we expect $k$ to show less variation with spin than $\eta_{\rm EM}$ or $l_{\rm EM}$. As an additional perk, fitting $k$ can inform us of the deviations from the standard BZ effect.

For $a=0$, there is no energy extraction along the field lines because $\Omega_{\rm H} = 0$, so we set the second term in equations (\ref{eq:dadt_diskjet}), (\ref{eq:dlnMdt_diskjet}), and (\ref{eq:spinup_diskjet}) to $0$. This might not be obvious in equations~\eqref{eq:dadt_diskjet} and \eqref{eq:spinup_diskjet}. However, because $\Omega_{\rm H} = ac/2r_{\rm H}$ and $\eta_{\rm EM} \propto a^2$, the $\Omega_{\rm H} \propto a$ term in the denominator cancels out: thus, for zero spin, this results in zero contribution from the jet terms (which contain $P_{\rm EM}$ and $\eta_{\rm EM}$).

With these constraints in our semi-analytic model we produce the thick red curve in Figure \ref{fig:spinup_param}, which gives us the total MAD spin-up $s_{\rm MAD}$ curve. This fits the data well, and we expect this semi-analytic model to be useful for modeling BH spin evolution. See Section~\ref{sec:discussion} for more details. 
The dashed lines in Fig.~\ref{fig:spinev}(b) show the spin evolution, due to the integration of Eqs.~\eqref{eq:dM} and \eqref{eq:dadlnM}, computed using a spline interpolation of the simulated data points, $e_{\rm in}(a)$ and $l_{\rm in}(M,a)$, shown in Fig.~\ref{fig:lin_ein}. In Fig.~\ref{fig:spinev}(b) we also show, in thin solid lines, the spin evolution computed using our semi-analytic model developed above. We notice a remarkable agreement between the spin-evolution computed using the interpolated data and the semi-analytic model.

There are some slight difference for $a_0=1$ that could be related to our model being less accurate around $a\sim0.5$. We stress that even with those slight differences our semi-analytic model reproduces the spin evolution with good accuracy and is adaptable for modeling BH spin in different accretion disk regimes and astrophysical contexts (e.g., Sec~\ref{sec:discussion_whodunnit}).

\begin{figure*}
    \centering
    \includegraphics[width=5in]{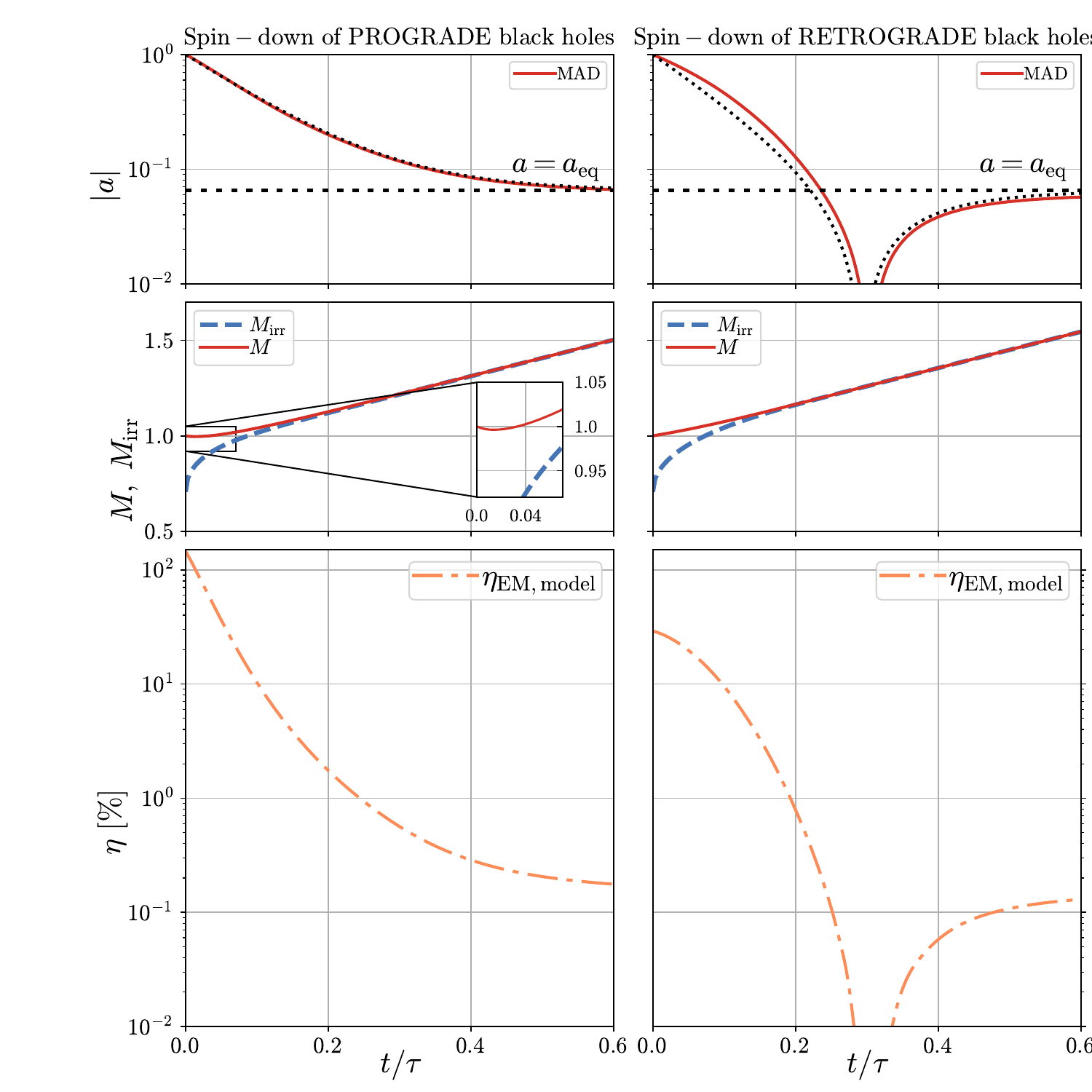}
    \caption{\text{Spindown of a maximally-spinning prograde BH down to the equilibrium spin results in the reduction of jet power by a factor of $\sim 10^3$.} BH spin, mass, and efficiency vs.{} time for  prograde (left column) and retrograde (right column) MADs with initial spins of $a_0=\pm1$, respectively. The time is measured in units of $\tau$, the timescale for the BH to accrete its own mass, $t/\tau = m/M_0$. [Top panel]: BH spin exponentially decays in time, closely following the fitting formula~\eqref{eq:a_vs_t_fit}. A MAD BH will spin down to $\aeq \approx 0.07$ by the time it has accreted a small fraction of its initial mass. For instance, a maximally spinning prograde BH spins down to $a = 0.2$ by $t/\tau = \boldsymbol{0.2}$, i.e., after consuming $\boldsymbol{20}\%$ of its own mass, and to $a=0.1$ by $t/\tau = \boldsymbol{0.35}$, i.e., after consuming $\boldsymbol{\simeq35\%}$ of its own mass. [Middle panel]: Total BH mass (red solid) and irreducible BH mass (blue dashed) vs. time. At $t/ \tau \lesssim \boldsymbol{0.04}$, the mass of the prograde BH briefly decreases before increasing again, as seen in the inset (left middle panel). This shows that BH loses rotational energy faster via the BZ mechanism than the accretion can replenish it (i.e., $\eta > 100\%$).  At later times, BH mass increases for both prograde and retrograde BHs. [Bottom panel]: Electromagnetic (EM) energy efficiency. In the prograde case, the EM efficiency drops by more than three orders of magnitude from $a=1$ to late times, when BH spin reaches its equilibrium value, $\aeq\approx 0.07$. Whereas for the integration in time to obtain the $a(t)$ dependence we use the interpolated values of $s$ measured from the simulations, we evaluate the EM efficiency, shown in the bottom panels, using the fits given in Fig.~\ref{fig:eta} and computed in Sec.~\ref{sec:results_MADmodel}.}
    \label{fig:retrograde}
\end{figure*}

In Figure \ref{fig:retrograde} we show the spin, mass and jet efficiency evolution in  MAD flows vs.\ time over a characteristic BH mass-doubling timescale $\tau$. The left hand column shows the prograde BH case with initial spin $a_0=1$, and the right hand column shows the retrograde case with $a_0=-1$. The timescale $\tau$ is defined such that $t/\tau = m/M_0$, where $m$ is the accreted mass and $M_0$ is the initial BH mass.  In the top panels of Fig.~\ref{fig:retrograde} we have fit the spin evolution to,
\begin{equation}
  a(t) = |a_0 - \aeq| e^{-t/\tau_{\rm MAD}} + \aeq,
  \label{eq:a_vs_t_fit}
\end{equation}
(black dotted line), where $\tau_{\rm MAD} = \tau/9.5$ and $a_0$ is the initial spin. 

The second row of panels from the top shows the dependence of BH mass $M$ vs.~$t$. The total BH mass, M is shown by the red lines, and the irreducible BH mass, $M_{\rm irr} = M (\frac{1}{2} r_{\rm H}(a))^{1/2}$, is shown  by the dashed  blue lines. The irreducible mass increases with time. The total BH mass increases with time for the retrograde BH, but for the prograde BH, we find that the total mass find that both the total mass and irreducible mass of the BH overall increase with time. While the BH is rapidly spinning, the total and irreducible BH masses are distinct because the BH also possesses rotational mass. As the BH spins down, $M_{\rm irr}$ and $M$ converge, after $t/\tau \approx 0.1$. 

The bottom panel shows the efficiency $\eta$ vs. mass-doubling time. In terms of the specific energy flux $e_{\rm in}$ onto the BH, we can write down the total efficiency with which the BH converts rest-mass energy into power:
\begin{equation}
    \eta_{\rm tot} = \frac{\dot m - \dot E}{\dot m} =  1 - e_{\rm in,tot}.
    \label{eta_1minusE} 
\end{equation} In terms of the hydrodynamic and electromagnetic specific energy fluxes, we can write it as
\begin{equation}
    \eta_{\rm tot} = \eta_{\rm HD} + \eta_{\rm EM} = 1 - e_{\rm HD} - e_{\rm EM}.
    \label{eta_tot} 
\end{equation} The energy fluxes $e_{\rm HD}$ and $e_{\rm EM}$ are computed using the fits derived above. As the BH spins down, the EM efficiency drops. In comparison, the hydrodynamic efficiency, $|1-e_{\rm HD}|$, remains roughly constant, consistent with the fact that in MADs, the disk torques do not significantly affect BH spin evolution with time.

Interestingly, the inset in the left middle panel of Fig.~\ref{fig:retrograde} shows that for the prograde case, the total BH mass slightly decreases before increasing. We see that during the short time the BH mass decreases, the rapidly rotating BH is spinning down so fast that the total efficiency $\eta$, the red curve in the bottom panel of Fig.~\ref{fig:retrograde}, exceeds $100\%$.

This is the first demonstration that, as a function of time, the BH actually \emph{loses} mass due to the jets outshining accretion. The cause of this phenomenon is that the BH rapidly loses rotation energy at high spin electromagnetically (through the jets), faster than the irreducible mass increases due to the energy supplied by accretion (through the disk).  

%\blcmt{}

\subsection{Analytic model to calculate $a_{\rm eq}$ in MADs}
\label{sec:results_analyticcheck}

Because the BH reaches $\aeq$ when $s=0$, we can work backward and use our semi-analytic model to find the value of \aeq{} analytically. We set equation (\ref{eq:spinup_diskjet}) to $0$ and solve for $\aeq$. As before, we use the average values in our model for the disk fluxes: $\langle l_{in, \rm HD}\rangle \approx 0.87$ and $\langle e_{in,\rm HD}\rangle \approx 0.97$. We use the positive fit $\eta_{EM} \approx 0.4 a^2$. For $k$ we use the fit for low positive spin, $k=0.1+0.5a$. For small $a$, $\Omega_{\rm H} \approx a/4$. Using these approximations, dropping any  terms that are higher order than $\propto a^2$ (they are small for low spin), we get a quadratic equation for \aeq{} that gives us the positive solution, $a_{\rm eq, MAD}^{\rm analytic}\approx 0.06$. This is remarkably close to the equilibrium spin value we have obtained directly from the simulation results, $a_{\rm eq, MAD}^{\rm simulated}\approx 0.07$ (see Sec~\ref{sec:results_spinev}).

\section{Discussion and Conclusion}
\label{sec:discussion}

\subsection{Semi-analytic Model of MAD Spin-down}
\label{sec:discussion_spindown-model}

We have used a suite of 3D GRMHD simulations of MADs from \cite{MAD_tchekho_2011,2012JPhCS.372a2040T,2015ASSL..414...45T} to study the BH spin evolution in MADs. \citet{2012JPhCS.372a2040T} were the first to find that a MAD will spin down a BH to an equilibrium spin of $a_{\rm eq}\approx 0.07$. Here, using the equations describing the BH spin evolution in time under the torques from the accretion disk and the jets, as measured in a suite of 3D GRMHD simulations, we showed that MADs will cause the BH to rapidly spin down, such that the BH reaches its equilibrium spin after accreting only $50 \%$ of its initial mass (see Fig.~\ref{fig:spinev}). Indeed, BH spin-down by a MAD disk is much more efficient than the BH spin-up by a NT disk (see Fig.~\ref{fig:spinev}).

Such a rapid spin-down might come as a surprise, because intuitively we think of accretion as spinning up the BH, by supplying it with the angular momentum. To understand the physics behind the MAD spin-down, we split up the torques acting on the BH into their hydrodynamic and magnetic components, as seen in Fig.~\ref{fig:spinup_param}. Figure~\ref{fig:Ang_Kep_drop} shows that some mechanism, likely magnetic fields threading the inner disk, robs the disk gas of its angular momentum before it can reach the BH: the MAD specific angular momentum lies below the Keplerian one and, in fact, it decreases towards the BH within the innermost stable circular orbit (ISCO), inside of which the specific angular momentum is assumed to be constant in the NT model. We thus end up with a strongly sub-Keplerian disk whose specific angular momentum decreases even further inside the ISCO and hence carries into the BH very little angular momentum, which is available for spinning up the BH. 

In addition to the magnetic fields extracting angular momentum from the BH gas supply, strong and dynamically-important magnetic fields, which thread the BH, extract the angular momentum from the BH and send it out to large distances in the form of jets. Thus, the low equilibrium BH spin is a consequence of both of these effects: the lower-than-expected angular momentum supplied by the disk gas pales in comparison with the large magnetically-mediated angular momentum extraction out of the BH by the large-scale magnetic flux (see Figs.~\ref{fig:lin_ein} and~\ref{fig:spinup_param}).

To quantify this, we have developed the first semi-analytic model that quantitatively describes what causes the BHs in the radiatively-inefficient MAD state to spin down so efficiently. We find that the hydrodynamic angular momentum and energy fluxes are essentially independent of the BH spin, as seen in Fig.~\ref{fig:lin_ein}. Thus, to the accuracy we are interested in, the hydrodynamic angular momentum ($l_{\rm HD}$) and energy ($e_{\rm HD}$) flux values -- which we take to be constants in our semi-analytic model (see Fig.~\ref{fig:lin_ein}) -- are unique properties of the MAD state, at least for the typical disk thickness that we have studied here ($h/r\approx0.3$). Using equation~\eqref{eq:spinup_diskjet}, we now can compute the hydrodynamic contribution to the spin-up parameter, $s_{\rm HD}$, which we plot with the orange dashed line in Fig.~\ref{fig:spinup_param}. It is remarkable that the gas retains so little angular momentum by the time it reaches the BH: in fact, $s_{\rm HD}$ vanishes at $a = 0.5$, which means that at $a > 0.5$, the hydrodynamic torques \emph{spin down} the BH. This means that if we were to consider a hypothetical hydrodynamic disk model with the structure identical to MAD (but without the magnetic fields), this model would result in $\aeq \simeq 0.5$.

The final step in building our semi-analytic model is to account for the EM torques, which we do via including the energy and angular momentum EM fluxes: we compute the values of these fluxes using the fitting formulas for the spin-dependence of the jet efficiency ($\eta_{\rm EM}$, Fig.~\ref{fig:eta}) and the angular velocity of the magnetic field lines ($k = \Omega_{\rm F}/\Omega_{\rm H}$, Fig.~\ref{fig:kvsa}).

\subsection{Whodunnit?}
\label{sec:discussion_whodunnit}

Equation~\eqref{eq:spinup_diskjet} gives the sum of the hydrodynamic and EM contributions, thus completing our semi-analytic model and giving us the spin-up parameter, $s$, which is shown in Fig.~\ref{fig:spinup_param} with the thick light red curve. The model is in good quantitative agreement with the numerical values for $s$, shown with dark red squares. This agreement is even more remarkable given that in order to keep the model as simple as possible, we purposefully used quite rough approximations for the free parameters in the model (e.g., $l_{\rm HD}$, $e_{\rm HD}$, and $k$), as we discussed in Sec.~\ref{sec:discussion_spindown-model}.

Figure~\ref{fig:spinup_param} shows that MADs have $|s_{\rm EM}| \gg |s_{\rm HD}|$, i.e., the EM torques dominate the hydrodynamic ones, over a wide range of spin, $a \gtrsim 0.2$. This is made uniquely possible by the strongly sub-NT angular momentum of MADs: the specific angular momentum of the MAD (orange dashed line) lies much lower than that of the NT disk (blue dash-dotted line).

To gauge the relative importance of the various processes for the rapid BH spin-down in MADs, we consider hypothetical disk models that differ in crucial aspects from our MAD simulations. For instance, in order to quantify the importance of sub-Keplerian nature of MADs on the spin-down, let us consider a disk model whose EM structure is that of the MAD, but whose hydrodynamic structure is that of the NT model. This combination might provide a rough description of luminous MADs, whose disks are thinner and hence closer to NT. For such disks, the equilibrium spin would shift to higher $a$ values to satisfy $s_{\rm NT} + s_{\rm EM} = 0$. From Fig.~\ref{fig:spinup_param}, we can approximately see that this happens at $\aeq \approx 0.3$. This means that even if the accretion flow were Keplerian (instead of significantly sub-Keplerian), the jets would still be strong enough to spin down the BH to a rather low, albeit higher, spin. If, additionally, the EM outflow efficiency drops substantially in luminous MADs compared to non-radiative ones (e.g., $\eta_{\rm EM}$ in Eq.~\ref{eq:spinup_diskjet} is lower by a factor of 2 than in our simulations, similar to what is seen in radiation-transport simulations of luminous MADs at $L/L_{\rm Edd} = 0.35$, \citealt{2022ApJ...935L...1L}), then the spin would evolve to $s_{\rm NT} + 0.5 s_{\rm EM} = 0$, resulting in the equilibrium spin of $\aeq \lesssim 0.5$.

The rather low equilibrium spins in the above hypothetical disk models can have interesting implications for luminous MADs, ones in which the large-scale magnetic flux is saturated on the BH, but those that can efficiently cool down to small values of scale height, $h/r \ll 1$. Such disks are particularly important because they can power luminous jetted quasars. The rotation profile of such thinner disks can be much closer to the Keplerian one than of our thick MADs. However, from the above arguments, it follows that even in this case the equilibrium spin can remain significantly below unity (e.g., $\aeq \sim 0.2-0.5$), defying the standard expectation of the ``standard'' NT disk model. 

\subsection{Comparison to other works}
\label{sec:discussion_otherworks}

There has been plentiful other theoretical work studying BH spin evolution. As we mentioned above, \cite{1970Natur.226...64B} was the first to compute an equilibrium spin for a BH surrounded by a turbulent disk, $a_{\rm eq}=1$. Fully turbulent relativistic disk solutions were also calculated by \cite{1998ApJ...504..419P}. They obtained smaller equilibrium spins, as low as $a_{\rm eq}=0.8$.
They achieved this by modifying the inner boundary condition for the angular momentum. They found that deviations from the Keplerian angular momentum profile controlled the equilibrium spin. Finally, the turbulent viscosity acting on the disk controls the angular momentum profile. We find striking similarities with our results as deviations from a Keplerian angular momentum profile are also important in our model.

\cite{2004ApJ...602..312G} were the first to use simulations of magnetized disks to measure the spin-up parameter and try to determine the equilibrium spin.
They computed a large equilibrium spin when compared to our simulations, $a_{\rm eq}\simeq0.93$. A possible explanation for this discrepancy is that their simulations are only 2D. Furthermore, compared to MADs, the jet power they measure in their simulations is weak \citep{mckinney_measurement_2004}. In this work, we find that the magnetized jet is quintessential for the rapid BH spin-down. Thus, we expect the difference in the equilibrium spin to be primarily the consequence of the difference in the jet power.

\cite{2022MNRAS.511.3795N} ran long-duration GRMHD simulations of MADs. They calculate the BH spin-up parameter as a function of BH spin. Using a fifth-degree polynomial fit for $a$, they measured an equilibrium spin of 0.04. This value is a factor of 2 smaller than ours. However, they only ran one simulation with a BH spin close to their \aeq, which may limit the accuracy to which $\aeq$ can be determined. The simulations of \cite{2022MNRAS.511.3795N} use an adiabatic index of $\Gamma=13/9$ and tested the effect of changing $\Gamma$ in their simulations and reported no significant difference in their results. We have run new simulations with the GPU-accelerated GRMHD code H-AMR \citep{2022ApJS..263...26L} with $\Gamma=13/9$ to study the effect of changing the adiabatic index. The spin-up parameter values for these simulations are shown by the black points in Figure \ref{fig:spinup_param}. The spin-up parameter values for all spins are shifted down from the $\Gamma=4/3$ simulations in red. These $\Gamma=13/9$ simulations are consistent with \cite{2022MNRAS.511.3795N}. Therefore, the difference in equilibrium spin may also be due to a different choice of $\Gamma$.

It is also possible that the spin-up parameter and the equilibrium spin weakly depend on time. Indeed, \cite{2022MNRAS.511.3795N} ran their simulations for a much longer time than ours. We investigated this possibility by computing the spin-up parameter as a function of time in our simulations. We found that the spin-up parameter does not show obvious trends in time, at least for the duration of our simulations.

In comparison to our work, \cite{2022MNRAS.511.3795N} focus on low-luminosity AGN and do not account for the change of BH mass when computing the time evolution of BH spin. This limits their ability to evolve the system to late times, when the BH spin approaches its equilibrium value.

\subsection{Observational implications}
\label{sec:discussion_observations}

Our 3D GRMHD simulations of radiatively-inefficient accretion flows can be applied to the super-Eddington accretion regime: in this regime, the accretion flow is so dense and optically-thick that the radiation does not have the time to escape out of the accretion flow and is instead advected into the BH or ejected as part of the outflow. Although radiatively-inefficient GRMHD simulations are usually interpreted in the context of very sub-Eddington accretion ($\dot M \ll 0.01 \dot{M}_{\rm Edd}$), we can make inferences for super-Eddington accretion by treating radiation-dominated optically-thick highly super-Eddington ($\dot M \gg \dot{M}_{\rm Edd}$) flows as an ideal  gas with the polytropic index, $\Gamma=4/3$, that corresponds to radiation-dominated gas. In an upcoming work (Lowell et al. 2024), we will show that the spin-down torques in this work are consistent with those in radiation transport simulations of super-Eddington BH flows.

Figure~\ref{fig:retrograde} shows the time evolution of the BH spin and the associated quantities over cosmological timescales. The bottom panels reveal that the spin-down leads to a considerable decrease in the jet efficiency, $\eta_{\rm EM}(a={1}) / \eta_{\rm EM}(a=\aeq) \simeq 10^3$. This can have significant observational consequences.

In the context of AGN, this could naturally explain the reported $10^3$ spread in the radio loudness of AGN \citep{2007ApJ_radioloud}. Indeed, radio-quiet quasars can have low spin, whereas radio-loud quasars can have high spin, as expected in the spin paradigm (Sec.~\ref{sec:intro}). The typical challenge facing this paradigm is that in order to explain such a large spread requires radio-quiet AGN to have very small spin, much lower than expected in the chaotic accretion model \citep{Volonteri_2007,2010ApJ...711...50T}. If AGN undergo spectral state transitions similar to microquasars, their central BHs would undergo periods of spin-down during the jetted MAD stages and spin-up during the jet-less stages. Thus, the radio-quiet AGN can form as the direct consequence of BH spin-down due to MADs. We note that other possibilities, such as the changes in the strength of the magnetic flux on the BH (relative to the MAD flux), changes in the ambient medium, or the presence of selection effects could contribute to the radio-loud/quiet dichotomy.

Gravitational wave detections \citep{abbott_observation_2016,abbott_gwtc-1_2019,abbott_gwtc-2_2021} have led to new ways of constraining BH spin. The gravitational wave signals contain information on the angular momentum of the BHs and their orbits. However, disentangling the BH spin distribution from this signal requires careful Bayesian analysis and is subject to multiple biases. Many teams have performed such type of analysis \citep{abbott_binary_2019,wysocki_reconstructing_2019,garcia-bellido_bayesian_2021,edelman_cover_2022,callister_no_2022,tong_population_2022,Ligo2022}.
They consistently find that the BH spin distribution is centered around small spins, $a_{\rm peak} < 0.3$ \citep[see however][]{galaudage_building_2021}. This distribution is consistent with our measurement of the equilibrium spin. 
Whereas it is unknown whether super-Eddington accretion or the MAD state are present in the formation and evolution LIGO/VIRGO/KARGA BHs, in certain BH formation scenarios it is possible to infer the presence of a MAD state and BH spindown. For instance, \citet{2023arXiv230207281J,2023ApJ...952L..32G} argue that BH spindown is inevitable in the context of collapsars, dying stars whose core collapses into a BH. Conversely, other evolutionary pathways, where the MAD state and super Eddington accretion are absent, could lead to a BH spin distribution centered around small spins. In this context, it is intriguing that BH spins measured in X-ray binaries and AGN have populations of both slowly and rapidly spinning BHs \citep{reynolds_observational_2021,fishbach_apples_2022}.

Gamma-ray bursts are some of the most powerful astrophysical phenomena in the universe. The collapsar model describes GRBs as the consequence of a failed supernova where all the energy of the gravitational collapse is funneled into a jet \citep{woosley_gamma-ray_1993}. The jet is fed by the rotation of a central engine, a BH, through the BZ mechanism.
Even though they are short events, lasting around 100 seconds, they are subject to very high accretion rates and are highly super-Eddington.

Recent large-scale GRMHD simulations of collapsars bridging the gap between the BH scales and the photosphere were run by \cite{gottlieb_black_2022a,gottlieb_black_2022b}. They showed that their simulations achieve the MAD state and launch powerful jets. They also found that, for realistic conditions, the accretion rate can reach 0.01 solar masses per second. This accretion rate would lead to around 1 solar mass accreted for a GRB lasting around 100 seconds. For a BH of 2 solar masses, this would lead to a considerable spin-down (see Fig.~\ref{fig:spinev}).  We should expect the remnant BHs to have a spin distribution centered around the equilibrium spin. The spin evolution during GRBs will be the subject of future work.

Furthermore, such a powerful spin-down could shut off the jet produced by the BH engine (see jet efficiency in Fig.~\ref{fig:retrograde}, and recall that ${m}/{M_0}={t}/{\tau}$). If the jet power is reduced considerably, the weaker jet could have trouble escaping the stellar envelope. However, the engine activity may be controlled by the magnetic reservoir in the stellar envelope instead of the angular momentum reservoir in the central BH \citep{gottlieb_black_2022a}. Thus, our results imply that collapsar models need to take into account the effects of BH spin-down to model the evolution of the jet engine.

We do not expect the accretion disk to accrete at super-Eddington rates most of the time. Therefore, it is important to determine how radiative effects influence the spin evolution of BHs. Cooler disks will be thinner and will have different properties when compared to thicker super-Eddington disks: they tend to more closely resemble a Keplerian disk as the scale height decreases. Cooling will likely modify the angular momentum profile and possibly the jet structure. Therefore, a thinner MAD will have a different equilibrium spin. However, as described in Sec. \ref{sec:discussion_whodunnit}, the equilibrium spin is still expected to be much lower than unity, even in the presence of radiative effects. Future work will focus on the effects of cooling on the spin evolution of MADs.  This will provide insight into the physics and observational signatures of luminous quasars accreting at sub-Eddington rates, $1-10\%$ of $L_{\rm Edd}$.

\begin{acknowledgements}
We thank Ryan Hickox and Zoheyr Doctor for helpful discussions. We thank the referee for their helpful comments on the manuscript.
  BL acknowledges support by a National Science Foundation Graduate Research Fellowship under Grant No. DGE-2234667. BL also acknowledges support by a Illinois Space Grant Consortium (ISGC) Graduate Fellowship supported by a National Aeronautics and Space Administration (NASA) grant awarded to the ISGC. JJ and AT acknowledge support by the NSF AST-2009884 and NASA 80NSSC21K1746 grants. AT also acknowledges support by NSF grants AST-2107839, AST-1815304, AST-1911080, OAC-2031997, and AST-2206471. Support for this work was also provided by the National Aeronautics and Space Administration through Chandra Award Number TM1-22005X issued by the Chandra X-ray Center, which is operated by the Smithsonian Astrophysical Observatory for and on behalf of the National Aeronautics Space Administration under contract NAS8-03060. We acknowledge support by the NSF through resources provided by NICS Kraken, where simulations were
carried out, NICS Nautilus and TACC Frontera \citep{stanzione2020frontera}, where data were analyzed, and NCSA
MSS and TACC Ranch, where data were backed up, under grants TG-AST100040 (TeraGrid), AST20011 (LRAC), and AST22011 (Pathways). 
\end{acknowledgements}

\appendix
\label{appendix}
\section{Energy and angular momentum fluxes for a Novikov-Thorne disk}
\label{sec:appendix_fluxes}

We use the following analytic equations from \citet{Moderski&SikoraBHev} and \citet{1970Natur.226...64B} calculate the fluxes on the BH for the standard Novikov-Thorne disk.

The marginally-stable radius, or the ISCO radius, is given by
\begin{equation}
    R_{ms} = 3 + Z_2 - [(3-Z_1)(3+Z1+2Z_2)]^{1/2},
\end{equation} where

\begin{equation}
    Z_1 = 1+(1-a^2)^{1/3} [(1+a)^{1/3} + (1-a)^{1/3}]
\end{equation} and

\begin{equation}
   Z_2 = (3a^2 + Z_1^2)^{1/2}.
\end{equation} Within the marginally-stable orbit, the energy and angular momentum fluxes are conserved. Using these, we calculate the specific energy flux on the BH horizon to be

\begin{equation}
   E_{\rm in} = (1-\frac{2}{3 R_{\rm ms}})^{1/2}.
\end{equation} The specific angular momentum flux on the BH is 
\begin{equation}
   L_{\rm in} = \frac{2M}{3^{3/2}} (1 + 2(3R_{\rm ms} - 2)^{1/2}).
\end{equation}

\section{Handling numerical floors}
\label{sec:appendix_floors}
This appendix describes how we avoid numerical floor contamination of the hydrodynamic fluxes computed in Section~\ref{sec:results}. In the integrals, Eqs.~(\ref{eq:fEoverfM}) and (\ref{eq:fLoverfM}), we only consider regions where $\sigma < 30$. This procedure excludes cells in the jet that are subject to numerical gas floors.

\begin{figure}
    \centering
    \includegraphics[width=0.5\columnwidth]{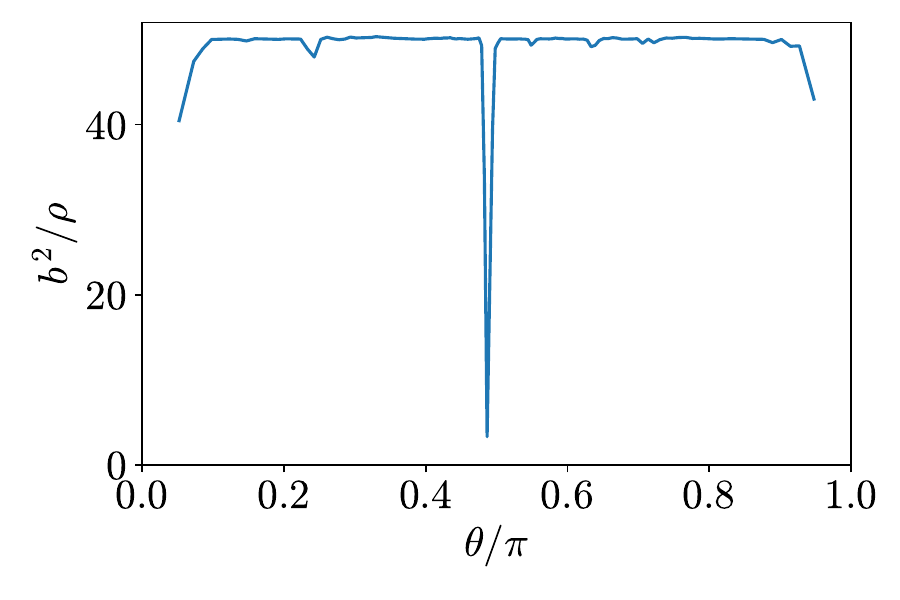}
    \caption{Example of an instantaneous polar profile of an azimuthal average of $b^2/\rho$, as evaluated at the BH event horizon, for $\sigma_{\rm  max} = 50$. At $\theta/\pi = 0.5$ we see the density dominated disk, whereas in the polar regions, $|\theta|/\pi\gtrsim 0.1$, we notice the contamination due to density floors. Due to the steepness of the profile a $\sigma$ cut, $\sigma<30$, allows to select most of the disk while removing the floor contamination in highly magnetized polar regions.}
    \label{fig:sigma_cut}
\end{figure}

Figure \ref{fig:sigma_cut} shows the $\sigma$ profile as a function of theta measured at $r=r_{\rm H}$ and azimuthally averaged. The gas is the dominant component in a thin sheet around $\theta=\pi/2$. Hence, we choose a cutoff of $\sigma=30$ that removes the floor contamination, visible in the polar regions, at $|\theta/\pi-0.5|\gtrsim 0.1$. Different snapshots in time show similar trends.

\begin{figure}
    \centering
    \includegraphics[width=0.5\columnwidth]{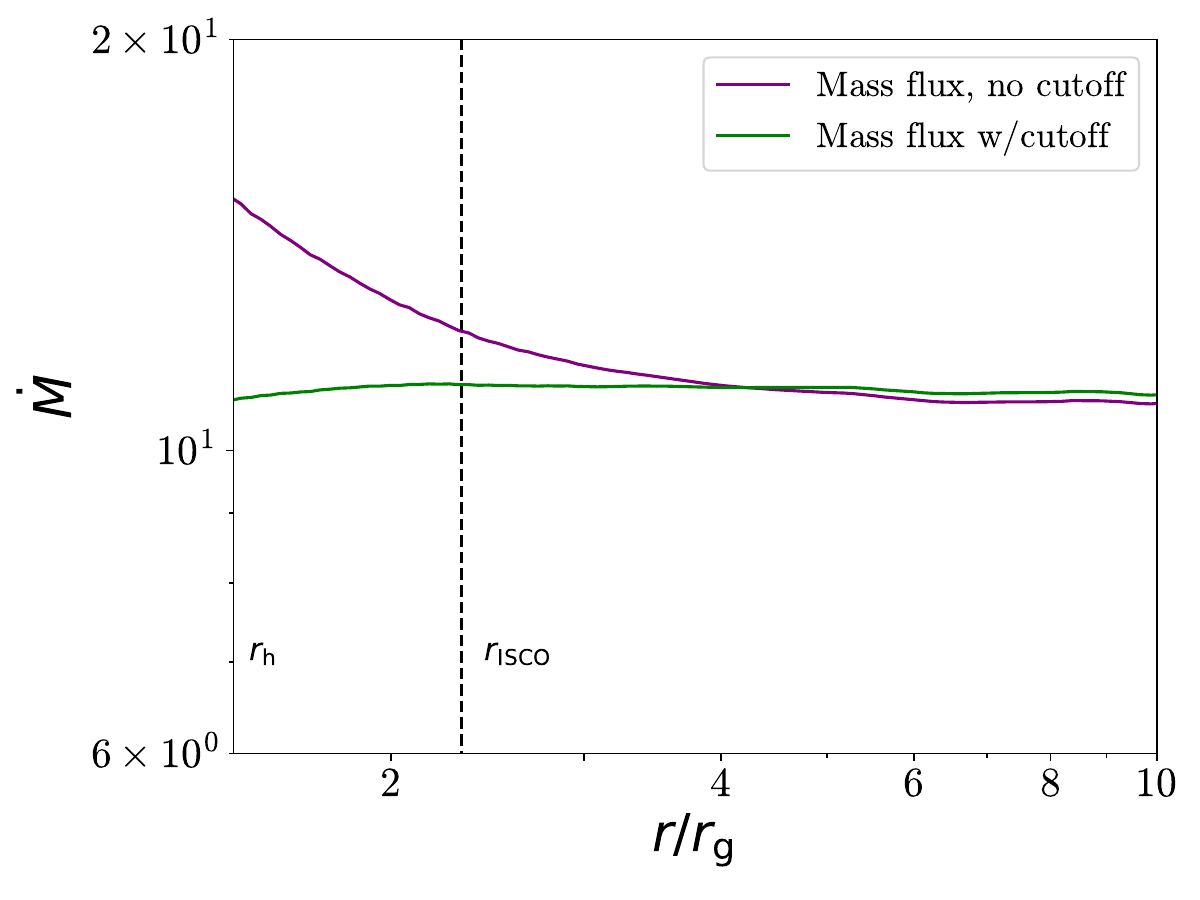}
    \caption{Time-average accretion rate as a function of radius, as computed using Eq.~(\ref{eq:fM}) for a simulation with $a=0.9$ with (green line) and without (purple line) using the $\sigma<30$ cutoff. Notice that once the cutoff is applied the accretion rate profile becomes flatter at the inner radii. This way, we can remove the density floor contamination from mass and energy accretion rates.}
    \label{fig:mdot_cutoff}
\end{figure}

In Fig.~\ref{fig:mdot_cutoff} we show the accretion rate (Eq.~\ref{eq:fM}) computed with and without the $\sigma$ cutoff. The accretion rate, corrected using our $\sigma$ cutoff, has a flatter profile at the inner radii when compared to the uncorrected one. We can remove the accretion boost due to the numerical floor contamination of density. 
We can then use this $\sigma$ cutoff to compute the fluxes for quantities that steeply depend on radii like $l_{\rm HD}$.  We apply the sigma cutoff to all hydrodynamic fluxes ($\dot{m}$, $l_{\rm HD}$, $e_{\rm HD}$). However, in practice $\dot{m}$ and $e_{\rm HD}$ are constant in radius after applying the $\sigma$ cutoff and could be safely measured at $5r_g$. 
The electromagnetic fluxes, $l_{\rm EM}$ and $e_{\rm EM}$, require no correction as they do not include quantities contaminated through the numerical floors (enthalpy, density, pressure).  

We have run an extra simulation with a different density floor value, given by a smaller $\sigma_{\rm max} =5$ instead of $\sigma_{\rm max}=50$ that is used throughout the manuscript. This factor of $10$ increase in the density floors, coupled with a lower cutoff value, $\sigma < 5$, did not lead to any significant differences in the measured values of the fluxes.

\bibliography{biblio.bib}

\end{document}